\documentclass[12pt,english]{article}
\usepackage{lmodern}
\usepackage[T1]{fontenc}

\usepackage{amsmath}
\usepackage{amssymb}
\usepackage{esint}

\usepackage[letterpaper,margin=1in,bottom=1in]{geometry}

\makeatletter
\usepackage{hyperref}
\usepackage{cite}
\usepackage{babel}
\usepackage{mathdots}
\usepackage{amsfonts}
\usepackage{mathrsfs}
\usepackage{amsmath}
\usepackage{esint}
\usepackage{graphicx}
\usepackage{youngtab}
\usepackage{multicol}
\usepackage{multirow}
\usepackage{slashed}
\usepackage{bm}
\usepackage{relsize}
\usepackage[dvipsnames]{xcolor}

\newcommand{\beq}{\begin {equation}}
\newcommand{\eeq}{\end   {equation}}
\newcommand{\bea}{\begin {eqnarray}}
\newcommand{\eea}{\end   {eqnarray}}
\newcommand{\baa}{\begin {array}   }
\newcommand{\eaa}{\end   {array}   }
\newcommand{\bit}{\begin {itemize} }
\newcommand{\eit}{\end   {itemize} }
\newcommand{\be }{\begin {equation}}
\newcommand{\ee }{\end   {equation}}
\newcommand{\nn }{\nonumber}

\allowdisplaybreaks

\makeatother

\usepackage{babel}

\begin{document}

\author{Wan-Zhe Feng$^a$\footnote{Email: vicf@tju.edu.cn}~~and Jiang-Hao Yu$^{b,c}$\footnote{Email: jhyu@itp.ac.cn}\\
$^{a}$ \textit{\small Center for Joint Quantum Studies and Department of Physics,}\\
\textit{\small School of Science, Tianjin University, Tianjin 300350, PR. China}\\
$^{b}$ \textit{\small Institute of Theoretical Physics, Chinese Academy of Sciences, Beijing 100190, PR. China}\\
$^{c}$ \textit{\small School of Physical Sciences, University of Chinese Academy of Sciences, Beijing 100049, PR. China}}

\title{\LARGE  \textbf{Twin Cogenesis}}

\date{}
\maketitle

\begin{abstract}
  We investigate a cogenesis mechanism within the twin Higgs setup which can naturally explain the nature of dark matter, the cosmic coincidence puzzle, little hierarchy problem, leptogenesis and the tiny neutrino masses. Three heavy Majorana neutrinos are introduced to the standard model sector and the twin sector respectively, which explain the tiny neutrino masses and generate the lepton asymmetry and the twin lepton asymmetry at the same time.
  The twin cogenesis mechanism applies to any viable twin Higgs model without an explicit $\mathbb{Z}_2$ breaking in the leptonic sector
  and evading the $\Delta N_{\rm eff}$ constraint.
  We illustrate the twin cogenesis mechanism using the neutrino-philic twin two Higgs doublet model, a newly proposed model to lift the twin neutrino masses with spontaneous $\mathbb{Z}_2$ breaking.
  The dark photon with a Stueckelberg mass $\mathcal{O}(10)$~MeV ensures the energy in the twin sector as well as the symmetric component of twin sector particles can be depleted. The lightest twin baryons are the dark matter candidates with masses approximately 5.5~GeV, which explain naturally the amount of dark matter and visible matter in the Universe are of the same order.
  We also demonstrate twin cogenesis in the fraternal twin Higgs setup,
  in which the dark matter candidate is the twin bottom bound state $\Omega^\prime_{b^\prime b^\prime b^\prime}$.
\end{abstract}

\newpage

\tableofcontents

\section{Introduction}\label{sec:Intro}

In this work, we discuss a mechanism which can naturally explain
the nature of dark matter, the cosmic coincidence puzzle,
little hierarchy problem, leptogenesis and the tiny neutrino masses.
In spite of the great success of the Standard Model (SM),
there are still number of remaining puzzles waiting to be solved.
First of all, the radiative corrections to the Higgs boson mass square
would inevitably make the Higgs mass huge, up to the Planck scale (or some intermediate scale)
unless there is an extreme fine-tuning cancellation,
this problem is referred to as the hierarchy problem (or little hierarchy problem).
Since the SM does not include right-handed neutrinos,
it cannot explain the tiny observed neutrino masses.
In addition, baryon (and lepton) asymmetry in the Universe is also among one of the unsolved puzzles,
usually presented by the observed baryon-to-photon ratio in the Universe~\cite{Komatsu:2010fb}
\begin{equation}
n_B / n_\gamma \sim 6\times 10^{-10}\,. \label{RBs}
\end{equation}
In the SM this ratio is computed to be too small to fit the observation,
which indicates the existence of new physics beyond the SM.
On the other hand, SM also does not provide a dark matter candidate to
explain the astrophysical observations.
Further one has the cosmic coincidence puzzle, i.e.,
the fact that the amount of dark matter and visible matter in the Universe are of the same order.
More specifically one has~\cite{Ade:2013sjv, Planck:2018vyg}
\begin{equation}
\frac{\Omega_{\rm{DM}}h^2_0}{\Omega_{\rm{B}}h^2_0}
=\frac{n_{\rm{DM}} m_{\rm{DM}}}{n_{\rm{B}} m_{\rm{B}}}
\approx 5.5 \,, \label{RDMB}
\end{equation}
which suggests dark matter and baryon asymmetry may have a common origin.
This can be explained by the so-called asymmetric dark matter hypothesis
that the dark particles are in thermal equilibrium with SM particles in the early universe,
and their chemical potentials and thus their number densities are of the same order.
Hence, if the mass of the dark particle is of the same order as the mass of baryons,
the cosmic coincidence Eq.~(\ref{RDMB}) can be naturally satisfied.
For a sample of asymmetric dark matter works see~\cite{Kaplan:2009ag,Cohen:2010kn,Ibe:2011hq,Feng:2012jn},
and for reviews see~\cite{Davoudiasl:2012uw,Petraki:2013wwa,Zurek:2013wia}.

The null result of new particle searches at the TeV scale from the Large Hadron Collider (LHC) challenges traditional solutions to the hierarchy problem, which usually predict new degree of freedom at the TeV scale.
The paradigm of neutral naturalness~\cite{Craig:2014aea, Chacko:2005pe, Burdman:2006tz, Cai:2008au, Csaki:2017jby, Serra:2017poj, Xu:2018ofw, Cohen:2018mgv, Xu:2019xuo, Ahmed:2020hiw} addresses this hierarchy problem by incorporating sub-TeV scale color neutral top partners,
which are less constrained at the LHC.
In particular, mirror twin Higgs model~\cite{Chacko:2005pe} is the prime and first example of neutral naturalness, which
introduces a mirror copy of SM particles, referred to as the twin sector, charged under the mirror SM gauge groups only.
The mirror symmetry can be traced back to the idea of
mirror world scenario~\cite{Lee:1956qn,Kobzarev:1966qya,Pavsic:1974rq,Blinnikov:1982eh}.
A discrete $\mathbb{Z}_2$ symmetry between the SM and the twin sector ensures that the quadratic divergences from the SM gauge bosons and top quark loop corrections on the Higgs mass square are cancelled by contributions from the mirror particles running in the loop. This protects the Higgs mass from radiative corrections up to some intermediate scale $\Lambda$, and thus solves the little hierarchy problem.
While this setup is theoretically appealing, it is difficult to reconcile with cosmological observations.
It predicts light twin neutrinos and twin photon giving too large contribution to the energy density of the Universe at late times, which has been excluded by the observations of the big bang nucleosynthesis (BBN) and anisotropies in the cosmic microwave background (CMB).

This cosmological tension can be alleviated by relaxing
the exact $\mathbb{Z}_2$ symmetry between the SM and the twin mass spectra.
The first proposal was the fraternal twin Higgs model~\cite{Craig:2015pha},
in which only the third generation of SM fermions was copied in the twin sector and the twin hypercharge was not gauged.
Other proposals include imposing hard breaking of the $\mathbb{Z}_2$ symmetry in Yukawa couplings of the twin sector~\cite{Barbieri:2016zxn,Barbieri:2017opf}, lifting the twin neutrino masses~\cite{Csaki:2017spo}, raising the mass of twin charged particles~\cite{Harigaya:2019shz}.
All of these solutions allow a great reduction of the number of degrees of freedom
in the twin sector at late times, leading to a suppression of the dark radiation.
Dark matter can be naturally accommodated within these proposals~\cite{Craig:2015xla, Garcia:2015loa, Garcia:2015toa, Farina:2015uea} and twin baryogenesis was also addressed~\cite{Farina:2016ndq} along this direction.

An alternative approach to resolve this tension is the asymmetric reheating mechanism~\cite{Chacko:2016hvu,Craig:2016lyx} in which the hard $\mathbb{Z}_2$ breaking is absent. In general, late asymmetric reheating can be realized without requiring further breaking of the discrete $\mathbb{Z}_2$ symmetry.
The late-time asymmetric reheating would dilute the twin energy density
by preferentially heating up the SM sector rather than the twin sector,
and thus evade the cosmological difficulty.
However, this dilution of twin energy density would cause the problem of the population of dark matter,
and washout any existing matter-antimatter asymmetry prior to the low temperature reheating process.
This presents a major challenge for such kind of $\mathbb{Z}_2$ symmetric solutions.
\cite{Koren:2019iuv} investigates a possible solution to this problem.

Inspired by the spontaneous $\mathbb{Z}_2$ symmetry breaking mechanism proposed in~\cite{Yu:2016bku, Yu:2016swa, Beauchesne:2015lva},
we propose a twin cogenesis mechanism in the neutrino-philic twin two Higgs doublet model,
in which the cosmological tension can be resolved.
At the same time, we also address the nature of dark matter and generation of matter-antimatter asymmetry.
We introduce three heavy right-handed Majorana neutrinos to both the SM sector and the twin sector respectively,
which explain the tiny neutrino masses via the seesaw mechanism.
Mass mixings at high scales cause all of the six heavy Majorana neutrinos to couple to both sectors.
The CP-violating decay of the Majorana fields
generates the lepton asymmetry in the SM sector
and the twin lepton asymmetry in the twin sector,
which is referred to as the ``twin cogenesis'' mechanism.
Cogenesis~\cite{An:2009vq,Davoudiasl:2010am,Gu:2010ft,Falkowski:2011xh,Blinov:2012hq,Feng:2013zda,Biswas:2018sib,Earl:2019wjw}
provides an intriguing scenario that
the asymmetric visible matter and dark matter are produced from the same source,
which automatically assigns similar number densities to baryons and asymmetric dark particles.
If the mass of the dark particle is around the mass of baryons,
with the comparable size of the baryon and asymmetric dark matter number densities,
the cosmic coincidence Eq.~(\ref{RDMB}) is naturally explained.

The twin cogenesis can be incorporated into
other twin Higgs models in which the cosmological tension is addressed.
We also illustrate twin cogenesis in the fraternal twin Higgs setup.
In this work, we describe the generic features of the twin cogenesis
and investigate the thermal history in details.
Since there is no interaction exchanging particle asymmetries between the SM and the twin sector,
after the out-of-equilibrium decay of the heavy Majorana fields,
the asymmetries generated in the SM and the twin sector then freeze inside each sector.
The (twin) lepton asymmetry subsequently transfers to the (twin) baryonic sector via (twin) sphaleron processes.
A massive dark photon with Stueckelberg mass around 10~MeV
kinetic mixing with the SM hypercharge,
ensures the energy in the twin sector can be depleted.
In the same way, the symmetric component of twin sector particles
can also annihilate to SM electron-positron pairs mediated by the dark photon,
which guarantees the lightest (asymmetric) twin baryons occupy the vast majority of dark matter.
The dark matter consists of twin protons and twin neutrons with masses approximately 5.5~GeV
explain naturally the amount of dark matter and visible matter in the Universe are of the same order.
Future dark matter direct detection experiments with better sensitivities have the possibility
to test the twin cogenesis mechanism with 5.5~GeV twin baryons as the dark matter.

Bound states of new non-Abelian forces, normally referred to as dark baryons and dark mesons,
are interesting dark matter candidates, and have been studied extensively in recent years.
Various frameworks which can cooperate these dark composite states were discussed in~\cite{Kribs:2009fy,Cline:2013zca,Dienes:2016vei,Lonsdale:2017mzg,Berryman:2017twh,Gresham:2017zqi,Mitridate:2017oky}.
Astrophysical constraints as well as different detection methods in searching for these types of dark matter were also investigated in~\cite{Gresham:2018anj,Chacko:2021vin}.
The mirror twin Higgs scenario provides a natural setting for asymmetric dark matter,
and the twin baryons are natural dark matter candidates~\cite{GarciaGarcia:2015pnn,Farina:2015uea,Farina:2016ndq,Freytsis:2016dgf,Barbieri:2016zxn,Barbieri:2017opf}.
These dark composite states bounded by new non-Abelian forces
can cause dark nucleosynthesis and other astrophysical effects~\cite{Krnjaic:2014xza,Detmold:2014qqa,Hardy:2014mqa,Gresham:2017cvl}.
These dark baryons can further form dark stars
and this subject has been studied most recently in~\cite{Curtin:2019lhm,Curtin:2019ngc,Howe:2021neq,Hippert:2021fch}.

The paper is organized as follows:
In Section~\ref{sec:MTH},
we briefly review the mirror twin Higgs model,
and discuss the thermal history of the SM sector and the twin sector under the twin cogenesis mechanism.
We also comment on dark radiation problem of the mirror twin Higgs model in general.
In Section~\ref{sec:T2H},
we introduce a neutrino-philic twin two Higgs doublet model
as an illustration of the twin cogenesis.
A detailed discussion of the twin cogenesis mechanism is carried out in Section~\ref{Sec:CogMTH}.
Twin cogenesis within the fraternal twin Higgs setup is discussed in Section~\ref{sec:CogFTH}.
In Section~\ref{sec:Pheno} we discuss the phenomenology related to the twin cogenesis.
Conclusions are given in Section~\ref{sec:Con}.

\section{Thermal history of mirror twin Higgs models}\label{sec:MTH}

We consider the extension of the SM with a twin Higgs sector~\cite{Chacko:2005pe}, which solves the little hierarchy problem
at scale $\Lambda \equiv 4\pi f \sim \mathcal{O}(10)$~TeV.
A $\mathbb{Z}_{2}$ symmetry
at high scale introduces a copy of SM particles, referring to as the twin sector, indicated by ``$\,^\prime\,$''.
The twin sector exhibits $SU(3)^{\prime}\times SU(2)^{\prime}\times U(1)^{\prime}$
gauge symmetries, with $g_{3}^{\prime}(\Lambda)\approx g_{3}(\Lambda)$,
$g_{2}^{\prime}(\Lambda)\approx g_{2}(\Lambda)$ and $g_{Y}^{\prime}(\Lambda)\approx g_{Y}(\Lambda)$
where $g_{3},g_{2},g_{Y}$ are the SM gauge couplings
respectively.
The twin-Higgs doublet $H'$ is introduced to be charged under $SU(2)^{\prime}\times U(1)^{\prime}$.
In the mirror twin Higgs setup, the matter content in the twin sector includes three
generations of twin-quarks $(Q_{iL}^{\prime},u_{iR}^{\prime},d_{iR}^{\prime})$,
and twin-leptons $(L_{i}^{\prime},e_{iR}^{\prime})$, where the index
$i=1,2,3$ indicates the generation. These twin fermions are only charged under mirror gauge symmetries.

In the Higgs sector, the scalar potential, containing the SM Higgs doublet $H$ and the twin Higgs doublet $H'$,
preserves an approximate global symmetry $SU(4)$ due to the $\mathbb{Z}_2$ symmetry:
\bea
	V_{\rm scalar} = - \mu^2 (H^2 + H'^2) + \lambda (H^2 + H'^2)^2 - \delta (|H|^4 + |H'|^4)\,.
\eea
Here $\delta$ parametrizes the small $SU(4)$-violating but $\mathbb{Z}_2$-preserving
radiative corrections from the gauge groups and the Yukawa interactions.
This radiative correction triggers a vacuum expectation value (VEV) for the twin Higgs boson with $\langle H' \rangle = f$, and
causes the global symmetry  breaking $SU(4) \to SU(3)$ with four Goldstone bosons.
Therefore, the $SU(4)$-symmetric Higgs fields can be re-parameterized nonlinearly as
\bea
	{\mathcal H} \equiv \left(\begin{array}{c} H \\ H' \end{array}\right)
    = \exp\left(\frac{\rm i}{f}\Pi\right)
       \left(\renewcommand{\arraystretch}{0.3}
            \begin{array}{c}
             ~ \\ ~ \\ ~ \\
             \bm{0}_{1 \times 2} \\
             ~ \\ ~ \\ ~ \\
                  0 \\ ~ \\
                  f \\ ~ \\
            \end{array}\right)
	= f\left(\renewcommand{\arraystretch}{0.25}
            \begin{array}{c}
             ~ \\
             \frac{{\rm i} \sin(\sqrt{\bm{h}^{\dagger}\bm{h}}/f)}{\sqrt{\bm{h}^{\dagger}\bm{h}}}  \bm{h}  \\
             ~ \\ ~ \\ ~ \\
                  0 \\ ~ \\
                  {\scriptstyle \cos( \sqrt{\bm{h}^{\dagger}\bm{h}}/{f} )} \\ ~ \\
            \end{array} \right)
	\simeq
    \left(\renewcommand{\arraystretch}{0.25}
            \begin{array}{c}
             ~ \\ ~ \\ ~ \\
             {\rm i} \bm{h} \\
             ~ \\ ~ \\ ~ \\ 
                  0 \\ ~ \\
            {\footnotesize f - \frac{1}{2  f}\bm{h}^{\dagger}\bm{h}} \\ ~ \\
            \end{array}\right)\,,
\eea
where $\Pi$ represents the Goldstone matrix, and $\bm{h}^T = (h^+ , h^0  )$
is identified to be the SM Higgs doublet with the vev $v$, in which the electroweak {\it vev} $v_{\rm ew} \equiv f \sin \theta$, where  $\theta = v/f$ and $\cos\theta = \sqrt{1- v_{\rm ew}^2/f^2}$.
The SM Higgs becomes a pseudo-Nambu-Goldstone boson, which protects its mass from radiative corrections.
This pseudo-Nambu-Goldstone boson should be consistent with the Higgs observed at the LHC,
and this places a bound on the twin Higgs {\it vev} $f/v\geq3$~\cite{Khachatryan:2016vau}.

To explain the seesaw origin of the neutrino masses and possibly realize the leptogenesis,
it is necessary to introduce the right-handed Majorana neutrinos, which has been discussed in~\cite{Barbieri:2016zxn, Chacko:2016hvu, Csaki:2017spo, Bishara:2018sgl} in the mirror twin Higgs setup.  
The simplest possibility is to introduce
three heavy Majorana neutrinos $N_{i}$
with complex Yukawa coupling constants $\lambda_{ij}$.
These three $N_{i}$ are responsible for the neutrino masses of both the SM and the twin sector.
Since the right-handed neutrinos $N_{i}$ are Majorana fermions not carrying any charge,
they have the following interactions which connect the SM and the twin sector,
\begin{equation}
-\mathcal{L} = M_{ij} {N}^c_{i} N_{j} + \lambda_{ij}\varepsilon_{\alpha\beta}\bar{N}_{i}P_{L}L_{j}^{\alpha}H^{\beta}+\lambda_{ij}\varepsilon_{\alpha\beta}\bar{N}_{i}P_{L}L_{j}^{\prime\alpha}H^{\prime\beta}+h.c.\,,
\label{AsyGen1N}
\end{equation}
where $\alpha,\beta$ are $SU(2)$ and $SU(2)^{\prime}$ indices.
However,
this simple setup always gives three zero neutrino masses due to the mass mixing,
and thus is not a realistic model.\footnote{
One could also introduce additional neutrino and twin neutrino mass mixing terms to avoid this problem,
for example the Higgs triplets~\cite{An:2009vq} were introduced in the setup other than the mirror twin Higgs models.}
In the fraternal twin Higgs model only the third generation twin fermions are introduced,
and thus only the third generation neutrinos are predicted to be massless using the above setup.
After choosing a basis that the heavy neutrino masses are diagonal,
the neutrino Yukawa interactions are written as
\begin{equation}
-\mathcal{L}_{\rm FTH} = M_{i} {N}^c_{i} N_{i} + \lambda_{ij}\varepsilon_{\alpha\beta}\bar{N}_{i}P_{L}L_{j}^{\alpha}H^{\beta}
+\lambda_{i3}\varepsilon_{\alpha\beta}\bar{N}_{i}P_{L}L_3^{\prime\alpha}H^{\prime\beta}+h.c.\,,
\label{AsyGenFTH}
\end{equation}
which predicts the inverse hierarchy neutrino masses and the tau neutrino has zero mass.

To avoid  the three zero neutrino masses due to Eq.~(\ref{AsyGen1N}),
we introduce three heavy right-handed Majorana neutrinos $N_{i}$ ($N'_{i}$)
to the SM (twin) sector, responsible for the neutrino masses of the SM (twin) sectors, respectively.
The $\mathbb{Z}_2$ symmetry does not forbid mixings between  $N_{i}$ and $N'_{i}$.
Thus we expect that both $N_{i}$ and $N'_{i}$ are responsible for leptogenesis in both the SM and the twin sector.
The Lagrangian contains the following interactions
\begin{equation}
-\mathcal{L}_{\rm Yuk}= M_{ij} {N}^c_{i} N_{j} + M_{ij} {N}^{\prime c}_{i} N'_{j} +  m_{ij} {N}^{c}_{i} N'_{j} +  \lambda_{ij}\varepsilon_{\alpha\beta}\bar{N}_{i}P_{L}L_{j}^{\alpha}H^{\beta}+\lambda_{ij}\varepsilon_{\alpha\beta}\bar{N'}_{i}P_{L}L_{j}^{\prime\alpha}H^{\prime\beta}+h.c.\,,
\label{AsyGen2N}
\end{equation}
Taking the mass hierarchy $M \gg m \gg \lambda f$,
the mass eigenstates and the corresponding masses are approximately given by
\bea
	N^\pm_{i} = \frac{1}{\sqrt{2}}(N_i \pm N^{\prime}_i)\,,
\qquad M^\pm_{ij} = M_{ij} \pm \frac{1}{2}\left(m_{ij} + m^T_{ij}\right)\,.
\eea
It is always possible to choose a basis for $N^\pm$ where the mass matrix $M^\pm$ is diagonal with three positive and real eigenvalues.
In term of mass eigenstates $N^\pm$ above the electroweak symmetry breaking scale, we obtain the Yukawa interactions
\begin{equation}
-\mathcal{L}_{\rm MTH}= M^\pm_{i} {N}^{\pm\,c}_{i} N^\pm_{i}
+  \lambda_{ij}\varepsilon_{\alpha\beta}\bar{N}^\pm_{i}P_{L}L_{j}^{\alpha}H^{\beta} \pm \lambda_{ij}\varepsilon_{\alpha\beta}\bar{N}^\pm_{i}P_{L}L_{j}^{\prime\alpha}H^{\prime\beta}+h.c.\,,
\label{AsyGenII}
\end{equation}
which looks like we introduce two copies of the three right-handed neutrinos in Eq.~\ref{AsyGen1N},
with slightly different masses $M^\pm_i$ and the opposite Yukawa coupling for $N^-_{i}$.
Hence, both copies of $N^\pm_{i}$ provide the cogenesis for the SM and the twin sector,
which will be explored in detail in Section~\ref{Sec:CogMTH}.

Integrating out the heavy states, we obtain the following low energy effective Lagrangian
\bea
	\mathcal{L}_{\rm eff}=  \frac{\lambda_{ik}\lambda_{kj}}{M_k}\left[  (\overline{L_{i}^c} H) (L_{j} H)
	+  (\overline{L_{i}^{\prime c}} H') (L'_{j} H')
	- \frac{(m_{\rm d})_k}{M_k} (\overline{L_{i}^c} H)(L'_{j} H')
	\right]	+ h.c.\,,
\eea
where the diagonal matrix $m_{\rm{d}} = \frac{1}{2}{\rm diag}\left\{(m_{ij} + m^T_{ij})\right\}$.
These terms generate the neutrino and twin neutrino masses, and also induce washout effects after the asymmetries are generated.
The neutrino and twin neutrino masses are obtained as follows
\bea
	m_{ij}  \simeq  \lambda_{ik} \left(M_k^{-1} - (m_{\rm d})_k M_k^{-2}\right) \lambda_{kj} v^2\,,
\qquad m^{\prime}_{ij} \simeq   \lambda_{ik} \left(M_k^{-1} + (m_{\rm d})_k M_k^{-2}\right) \lambda_{kj} f^2\,.
\label{MTHnM}
\eea
Thus the twin neutrino masses are $\sim f^2/v^2$ times larger than the SM neutrino masses.
Therefore, these twin neutrinos will be kept in the hidden sector cosmology as dark radiation,
which brings cosmological problems.
Solutions to the dark radiation problem for twin Higgs models will be discussed at the end of this section.

We now investigate the thermal history of the twin sector in the early universe.
Since all the twin particles are not charged under the SM gauge groups,
the twin sector offers dark matter candidate naturally.
With the inclusion of the three heavy right-handed neutrinos,
both the dark matter candidate and the matter-antimatter asymmetry,
can be addressed at the same time.
Once the SM particles and the twin sector particles are created in the reheating phase after the inflation,
the SM and the twin sector particles are in thermal equilibrium with the exchange of the
six Majorana fermions $N^\pm_i$.
When the temperature drops down,
the six $N^\pm_i$ decay out-of-equilibrium,
generating a net lepton number in the SM sector
and also a net twin lepton number in the twin sector,
as described in Section~\ref{Sec:CogMTH}.
Now the SM sector and the twin sector are still
thermally connected by the Higgs mediated interactions,
such as $(H H^\dagger)(\bar{f}^\prime f^\prime)$, $(H^\prime H^{\prime \dagger})(\bar f f)$
and $(\bar f f)(\bar{f}^\prime f^\prime)$,
where $f,f^\prime$ denote SM and twin sector fermions.
However,
these interactions will not exchange the asymmetries generated in the SM and the twin sector.
Thus after the decay of $N^\pm_i$,
the asymmetries generated
will then freeze inside the two sectors.
Due to the (twin) sphaleron processes above the (twin) electroweak scale,
fraction of (twin) lepton number subsequently convert to the (twin) baryon number,
inducing an asymmetry in the (mirror) SM sector.

As the temperature drops down to around ${\mathcal O}$(GeV),
the thermal decoupling between the SM and the twin sector
occurs once the annihilation and scattering rates mediated by the Higgs boson fall below the Hubble rate.
After the thermal decoupling at temperature $T_d$, the SM and the twin sector will then evolute independently, with different temperatures $T$ and $T'$ respectively.
Assuming separate entropy conservation in both sectors, the temperature of the twin sector can be determined by
\bea
	\frac{T'}{T} = \frac{T'_d}{T_d}\left(\frac{g_*(T)}{g_*(T_d)}\right)^{1/3} \left(\frac{g'_*(T_d)}{g'_*(T)}\right)^{1/3}\,, \quad \textrm{with} \,\, T < T'_d = T_d \simeq {\mathcal O}(\textrm{GeV})\,,
\eea
where $g_*(T)$ ($g'_*(T)$) is the effective number of degrees of freedom,
summed over all relativistic degrees of freedom with weight 1 for a boson and 7/8 for a fermion
at a given temperature $T$, in the SM (twin) sector.
The $T'$ typically depends on both the decoupling temperature and the mass spectra in the twin sector.
Possessing relatively higher mass spectra,
the twin sector typically has lower temperature than the SM.
If the two sector decouple after the twin QCD phase transition,
but before the ordinary QCD phase transition,
the SM QCD phase transition is transferred to only the SM thermal bath,
while the entropy of the twin QCD phase transition is released
before the two sector thermally decoupled,
which makes the $T'$ much lower than all other cases.
After the twin quark-hadron phase transition,
the twin QCD has chiral symmetry breaking and all free twin quarks are confined into twin hadrons.

At this temperature, the light twin particles include twin hadrons, twin charged leptons, twin photon, twin neutrinos. Eventually the entropy of light twin particles is transferred into twin photons and twin neutrinos, which behave as extra radiation components. This is quantified through an effective neutrino number defined by
\bea
	N_{\rm eff} = 3 \left(\frac{11}{4}\right)^{4/3}\left(\frac{T_\nu}{T}\right)^{4} + \frac{4}{7} \left(\frac{11}{4}\right)^{4/3} g'_*\left(\frac{T'}{T}\right)^{4}\,.
\eea
Recent measurements on the Planck data gives the bound $N_{\rm eff} = 3.05 \pm 0.27$ at the 95\% confidence level~\cite{Planck:2018vyg}.

One of the major issues of the twin Higgs model is the additional number of
relativistic species (twin photon and twin neutrinos) might give too much contribution to
$\Delta N_{\rm eff}$.
A generic twin Higgs model contribute too large $\Delta N_{\rm eff}$
to agree with BBN and CMB observations~\cite{Planck:2018vyg},
mostly coming from the three twin neutrinos.
This tension can be solved in several different ways:
\bit
\item The simplest solution is removing the degrees of freedom of twin neutrinos,
such as making the twin quarks vector-like~\cite{Craig:2016kue},
or the fraternal twin Higgs setup~\cite{Craig:2015pha}
in which the twin sector only contains one generation of twin fermion
and a single twin neutrino with  $\Delta N_{\rm eff} \approx 0.075$ is consistent with the current observation.
In addition,
the twin photon can also be made massive to further reduce the dark radiation~\cite{Bishara:2018sgl,Batell:2019ptb,Liu:2019ixm}.
\item  Another simple solution is to raise the mass of the three twin left-handed neutrinos,
such that their contribution to $\Delta N_{\rm eff}$ can be removed.
This can be done by lifting the twin sector Yukawa couplings~\cite{Barbieri:2016zxn,Barbieri:2017opf}
or by assigning different Majorana masses to the right-handed neutrinos from the SM and the twin sector~\cite{Csaki:2017spo}.
\item In contrast to lifting the mass of twin neutrinos,
raising the mass of twin charged particles
allows twin neutrinos to decouple from the thermal bath at a much earlier time, about a few GeVs,
and thus the twin neutrinos contribution to $\Delta N_{\rm eff}$
can be diluted by other particles leaving the thermal bath afterwards~\cite{Harigaya:2019shz}.
\item Alternatively, the asymmetric reheating provide another solution to this problem.
A late decay of some additional particles dominantly to SM particles
after the the twin sector decouples from the SM sector
may dilute the energy density of the twin sector~\cite{Chacko:2016hvu,Craig:2016lyx}.
\eit
All the above solutions, except the last one,
need an explicit $\mathbb{Z}_2$ symmetry breaking in the fermionic sector.
Since the asymmetric entropy release must happen below the two sector decouples,
it is difficult to incorporate the high scale leptogenesis in this mechanism.

Twin cogenesis mechanism can apply to twin Higgs models
which can avoid too much dark radiation.
As illustrations,
in this work we will show twin cogenesis in two simple setups:
\bit
\item
We propose a new model in which a spontaneous $\mathbb{Z}_2$ symmetry breaking
lifts the twin neutrino masses above $\mathcal{O}$(MeV),
and thus has no contribution to the dark radiation.
\item
We realize the twin cogenesis in the fraternal twin Higgs setup
with only one species of light twin tau neutrino
which is consistent with the current limit on $\Delta N_{\rm eff}$.
\eit

However, just avoiding the additional contribution to $\Delta N_{\rm eff}$ is not enough
to provide a valid twin cogenesis.
The light twin states, e.g., twin electrons, or twin taus for the fraternal twin Higgs setup,
will not annihilate and thus would overclose the Universe.
Even for twin neutrinos with lifted masses around a few MeVs,
although no longer contributing to $\Delta N_{\rm eff}$,
they are stable and will be the dark matter candidate
contributing too much to the relic density.
Thus one needs to have a mechanism to dilute the energy in the twin sector.
This can be achieved by a massive twin photon $\gamma^{\prime}$ kinetic mixing with the SM hypercharge,
which will mediate the annihilation of light twin states to SM particles.
To this end, we introduce a tiny Stueckelberg mass around 10~MeV
to the twin photon $\gamma^{\prime}$ 
such that the it can decay into SM electron-positron pairs.
The detail of mixing will be discussed in Section~\ref{sec:Pheno}.

In the new model we proposed in Section~\ref{sec:T2H},
the twin neutrino masses are lifted to above $\mathcal{O}(10)$~MeV,
and the mass hierarchy in the twin sector can be written as
\begin{equation}
2 m_{f^\prime} > 2 m_{\nu_i^\prime} >  2 m_{e^\prime} > m_{\gamma^\prime} > 2 m_{e}\,,
\end{equation}
where $f^\prime$ denotes all other twin $U(1)^\prime$ charged fermions expect the twin electron $e^\prime$.
As the temperature drops down,
twin muons and twin taus will decay to twin electrons plus twin neutrinos through twin weak interactions.
The second and third generation twin quarks will also decay to twin up and down quarks through twin flavor mixings.
In addition, twin neutrinos with masses $\mathcal{O}(10)$~MeV or higher,
as well as all other heavier twin fermions,
can also annihilate to twin electrons through twin electroweak interactions.
In addition, twin gauge bosons will also decay to light twin sector fermions
and eventually annihilate to SM electron-positron pairs.
At the same time, twin $U(1)_{\rm EM}^\prime$ charged fermions
also annihilate to dark photons via twin electromagnetic interactions.
With the presence of kinetic mixing $\epsilon \sim 10^{-9}$ between SM and twin hypercharges,
the decay width of dark photon to SM electron-positron pair is given by $\sim \epsilon^2 \alpha m_{\gamma^\prime} /3$,
where $\alpha$ is the fine structure constant,
providing the $\mathcal{O}(10)$~MeV dark photons decay to SM electron-positron pairs
less than $10^{-3}$ second, consistent with the BBN constraint.
Thus the symmetric component of twin particles can also be depleted efficiently.
Therefore, all heavier twin particles as well as
the symmetric component of twin fermions are removed from the twin sector,
the dark matter candidates for twin cogenesis are the (asymmetric) twin baryons,
which explains the nature of dark matter
and the cosmic coincidence puzzle with
the ratio of dark baryon mass to the proton mass to be approximately 5.5.
The twin baryon masses are linked to the twin $SU(3)^\prime$ confinement scale.
Thus the twin cogenesis requires the twin QCD confinement scale
\begin{equation}
\Lambda^\prime_{\rm QCD} \approx 5.5 \,\Lambda_{\rm QCD}\,,
\end{equation}
which can be achieved in two different ways:
\begin{itemize}
\item
  Introducing a small difference between $\alpha_s^\prime(\Lambda)$ and $\alpha_s(\Lambda)$
  at the scale $\Lambda = 4\pi f$.
  $\Lambda^\prime_{\rm QCD} \approx 5.5 \,\Lambda_{\rm QCD}$ corresponds to
  an $\mathcal{O}(10\%)$ splitting of the QCD and twin QCD coupling constants through RGE running from the $\Lambda$.~\cite{Farina:2015uea}.
  This can be done via dynamics from scales higher than  $\Lambda$~\cite{Geller:2014kta,Csaki:2015gfd}.
\item
  Introducing additional $\mathbb{Z}_2$ breaking term to modify the twin $b^\prime$ quark Yukawa coupling, possibly originated from the mixing between the twin $b^\prime$ and heavy composite fermions~\cite{Barbieri:2015lqa}.  It has been shown in~\cite{Ahmed:2017psb} that a $\mathcal{O}(10\%-20\%)$ difference of $b$ and $b^\prime$ Yukawa couplings at $\Lambda = 4\pi f$ can result in
  $\Lambda^\prime_{\rm QCD}/\Lambda_{\rm QCD}\sim 5.5$.
\end{itemize}

In the fraternal twin Higgs setup,
twin light states as well as the symmetric component of twin fermions
will also annihilate to the SM electron-positron pairs through the dark photon
with mass $\mathcal{O}(10)$~MeV or higher.
While in this setup, the twin tau neutrino is light and will become dark radiation
but still within the current limit for $\Delta N_{\rm eff}$.
Dark matter candidates in the fraternal twin Higgs setup
will be explored in details in Section~\ref{sec:CogFTH}.

\section{Neutrino-philic twin two Higgs doublet model}\label{sec:T2H}

In this section,
we propose a new model that a spontaneous $\mathbb{Z}_2$ symmetry breaking
increases the twin neutrino masses to above $\mathcal{O}(10)$~MeV,
such that the twin neutrinos can annihilate to twin electron-positron pairs and
thus contribute to neither dark radiation nor dark matter relic density.
To this end,
we extend the minimal twin Higgs model to the twin neutrino-philic two Higgs doublet model (twin $\nu$2HDM).
In the visible sector, two Higgs doublets $H_{1}$ and $H_{2}$ are introduced as the 2HDM setup.
While in the twin sector, two twin Higgs doublets $H_{1}^\prime$ and $H_{2}^\prime$ are introduced,
mapped into the 2HDM Higgses via the twin parity:
$H_{1}^\prime \xrightarrow{\mathbb{Z}_2} H_{1}$, $H_{2}^\prime \xrightarrow{\mathbb{Z}_2} H_{2}$.
Imposing the $\mathbb{Z}_2$ symmetry,
the general tree-level scalar potential with
an approximate global $SU(4)$ symmetry reads
\bea
	V_{\rm tree} =
	-\mu_1^2 |H_1|^2 - \mu_2^2 |H_2|^2
	+ \lambda_1 (|H_1|^2)^2 + \lambda_2 (|H_2|^2)^2
	+ \lambda_3 |H_1|^2 |H_2|^2 +  m_{12}^2 \left[ H_1^\dagger H_2 +h.c.\right],
	\label{eq:2HDMPot}
\eea
where $m_{12}^2$ parametrizes the soft $SU(4)$ breaking.
The tree-level potential breaks the $SU(4)$ symmetry and one obtains the following Goldstone bosons:
\bea
{\mathcal H}_i \equiv \left(\begin{array}{c} H_i \\ H'_i \end{array}\right)
\simeq
\left(\renewcommand{\arraystretch}{0.3}
            \begin{array}{c}
             ~ \\
            {\rm i} \bm{h}_i   \\
             ~ \\ ~ \\ ~ \\
            {\rm i} C_i \\ ~ \\
            f_i - \frac{1}{2  f}\bm{h}_i^{\dag}\bm{h}_i + {\rm i} N_i  \\ ~ \\
            \end{array} \right)\,,
\quad i = 1, 2\,,
\eea
where one combination of the $C_i$ and $N_i$ are absorbed by the longitudinal components of the twin $W$ and twin $Z$ bosons, and
the Goldstone bosons $\bm{h}_i$ ($i=1,2$) are identified as the two Higgs doublets.
Details can be found in~\cite{Yu:2016bku, Yu:2016swa, Yu:2016cdr}. 

The fermion assignment in the twin $\nu$2HDM is as follows: the neutrinos and twin neutrinos only interact with the second Higgs doublet $H_2$ and $H'_2$:
\begin{equation}
- {\mathcal L}_{\rm Yuk} = M_{ij} {N}^c_{i} N_{j} + M_{ij} {N}^{\prime c}_{i} N'_{j} +  m_{ij} {N}^{c}_{i} N'_{j} +  \lambda_{ij}\varepsilon_{\alpha\beta}\bar{N}_{i}P_{L}L_{j}^{\alpha}H_2^{\beta}+\lambda_{ij}\varepsilon_{\alpha\beta}\bar{N'}_{i}P_{L}L_{j}^{\prime\alpha}H_2^{\prime\beta}+h.c.\,,\label{v2HDMYuk}
\end{equation}
while all other SM fermions and twin fermions interact with the SM Higgs doublet $H_1$ and $H'_1$.
Given this fermion assignment, the radiative corrections on the scalar potential are computed to be:
\begin{align}
	V_{\rm loop} &= \delta_1 \left(|H_{1}|^4 +  |H'_{1}|^4\right)
	+ \delta_2 \left(|H_{2}|^4 +  |H'_{2}|^4\right)
	+ \delta_3 \left(|H_{1}|^2 |H_{2}|^2 +  |H'_{1}|^2|H'_{2}|^2\right) \nn\\
	& + \delta_4 \left(|H_{1}^\dagger H_{2}|^2   +  |H^{'\dagger}_{1} H'_{2}|^2 \right)
	+ \frac{\delta_5}{2} \left[(H_{1}^\dagger H_{2})^2   +  (H^{'\dagger}_1 H'_{2})^2 + h.c. \right],
	\label{eq:loopcorr}
\end{align}
where $\delta_i$ ($i=1,\cdots, 5$) parameterize the radiative corrections from the gauge and Yukawa interactions.
For detailed expressions, we refer to \cite{Yu:2016bku, Yu:2016swa, Yu:2016cdr}.

With all the terms of the scalar potential above, we obtain the {\it vev}s of the two Higgs doublets
\bea
	\langle {\mathcal H}_1 \rangle \equiv \left(\begin{array}{c} 0 \\ f_1 \sin \theta_1 \\ 0 \\ f_1 \cos \theta_1 \end{array}\right), \qquad
	\langle {\mathcal H}_2 \rangle \equiv \left(\begin{array}{c} 0 \\ f_2 \sin \theta_2 \\ 0 \\ f_2 \cos \theta_2 \end{array}\right).
\eea
with
\bea
	\theta_1 \equiv \frac{\langle h_{1} \rangle}{f_1}\,, \quad \theta_2 \equiv \frac{\langle h_{2} \rangle}{f_2}\,.
\eea
From the full potential, the tadpole conditions read
\bea
	 && \delta_1  \sin 4\theta_1 + \delta_2\tan^4\beta  \sin 4 \theta_2  +  \delta_{345} \tan^2\beta   \sin 2(\theta_1 + \theta_2) = 0\,,\\
	 && \delta_1  \sin 4\theta_1 - \delta_2\tan^4\beta  \sin 4 \theta_2  +  \delta_{345} \tan^2\beta  \sin 2(\theta_1 - \theta_2)
	  -    \frac{4m_{12}^2}{f_1^2} \tan\beta \sin(\theta_1 - \theta_2) = 0\,. \nn
	\label{eq:TadpoleU4U4}
\eea
According to \cite{Yu:2016swa}, if $m_{12}^2$ is zero, radiative $\mathbb{Z}_2$ breaking induces $\langle h_{1} \rangle = v_{\rm ew}$ and $\langle h_{2} \rangle = 0$; while
if $m_{12}^2$ is gradually turned on, the tadpole induced $\mathbb{Z}_2$ breaking induces $\langle h_{1} \rangle \simeq v_{\rm ew}$ and $\langle h_{2} \rangle = u$.
Here we take small $m_{12}^2$ parameter and realize the hierarchical {\it vev}s $f_2 \gg f_1 \gg v_{\rm ew} \gg u$, e.g.,
$\langle h_{2} \rangle \sim 0.001$ GeV while $f_2 \sim 2$ TeV.
We note this hierarchical {\it vev} structure does not affect the fine tuning level in the twin Higgs model:
\bea
	\Delta_m = \left|\frac{2\delta m^2}{m_h^2}\right|^{-1}  \simeq   \frac{2v^2}{f_1^2}\,.
\eea
If we keep the scale $f_1$ to be $\mathcal{O}({\rm TeV})$, this corresponds to around $15$\% tuning.
According to Eq.~\ref{MTHnM}, the neutrino masses and twin neutrino masses are obtained as follows
\bea
	m_{ij}  \simeq  \lambda_{ik} \left(M_k^{-1} - m_{{\rm d} k} M_k^{-2}\right) \lambda_{kj} u^2\,,
\qquad m^{\prime}_{ij} \simeq   \lambda_{ik} \left(M_k^{-1} + m_{{\rm d} k} M_k^{-2}\right) \lambda_{kj} f_2^2\,.
\label{MTHnM}
\eea
In this case, the twin neutrino masses are proportional to the SM neutrino masses with the relation
\bea
	m^{\prime}_{ij}  \simeq \frac{f_2^2}{u^2}\,m_{ij}\,.
\eea
So taking $u \sim 0.1$ GeV while $f_2 \sim 2$ TeV, and assuming nearly degenerate neutrino masses,
we can lift the twin neutrino masses to the MeV scale or higher.
Therefore, this model provides a new way to increase the twin neutrino masses to avoid $\Delta N_{\rm eff}$ constraints.

\section{Twin cogenesis for mirror twin Higgs models}\label{Sec:CogMTH}

Twin cogenesis mechanism also applies to other twin Higgs model
in which the Yukawa couplings of twin matter fermions are identical with the corresponding SM fermions.
Without specifying a concrete twin Higgs model,
in this section we will discuss in general the twin cogenesis mechanism for mirror twin Higgs setup
with three generations of twin fermions,
and we assume the problem of too much dark radiation due to twin neutrinos can be addressed in such models,
e.g., the $\nu$T2HDM proposed in Section~\ref{sec:T2H}.

Aside from explaining the tiny neutrino masses,
the six heavy right-handed Majorana neutrinos $N^{\pm}_i$
can also generate asymmetries in both the SM and the twin sector,
via interactions Eq.~(\ref{AsyGenII}).
The CP violations due to the complex coupling $\lambda_{ij}$ generate
the excess of $L_{i}^{\alpha}H^{\beta}$ over $\bar{L}_{i}^{\alpha}H^{\beta*}$
and $L_{i}^{\prime\alpha}H^{\prime\beta}$ over $\bar{L}_{i}^{\prime\alpha}H^{\prime\beta*}$,
and thus create a net $B-L$ number in the SM sector
and create a net $B^\prime-L^\prime$ number in the twin sector.
Since the complex coupling $\lambda_{ij}$ are identical for the SM and the twin sector,
the interactions Eq. (\ref{AsyGenII}) generate (almost)
the same amount of asymmetry in both sectors.
Hence, the $B^\prime-L^\prime$ number created in the twin sector
is equal to the $B - L$ number created in the SM sector
(if the washout effects are similar in both the SM and the twin sector), i.e.,
\begin{equation}
B^{\prime}-L^{\prime} \approx (B-L)\,.\label{TBL}
\end{equation}
After the out-of-equilibrium decay of the heavy Majorana fields $N^\pm_i$,
these asymmetries freeze in both sectors, since there is no interaction
transferring the $B-L$ or $B^\prime-L^\prime$ excess between the SM and the twin sector.
As mentioned in Section~\ref{sec:MTH},
although interactions between the two sectors are still present,
interactions like $(H H^\dagger)(\bar{f}^\prime f^\prime)$, $(H^\prime H^{\prime \dagger})(\bar f f)$
and $(\bar f f)(\bar{f}^\prime f^\prime)$
where $f,f^\prime$ denote SM and twin sector fermions,
will not exchange the asymmetries generated in the SM and the twin sector.
The asymmetries generated via the  $N^\pm_{i}$  decay would then pass to (twin) baryons through (twin) sphaleron processes.
Once the asymmetry was generated, the total $B-L$ number is conserved in both the SM and the twin sector
since the sphaleron processes break $B+L$ while preserve $B-L$.

We now turn to the details of the genesis of lepton number asymmetry
in the SM sector as well as the twin sector.
The lepton asymmetry is generated from the
CP violation decays of the Majorana field $N^\pm_i$
into the left-handed lepton doublet $L_{j}$ and the Higgs doublet $H$.
The asymmetry is created by the interference of the tree and loop diagrams,
which consist of a vertex diagram and a wave function diagram~\cite{Covi:1996wh}.
Assuming $M_{2,3}\gg M_1$ in Eq.~(\ref{AsyGenII}),
the asymmetry is generated mostly from the decay of $N^\pm_1$.
The excess of $L_{j}^{\alpha}H^{\beta}$
over $\bar{L}_{j}^{\alpha}H^{\beta*}$ is given by~\cite{Covi:1996wh,Feng:2013zda}
\begin{align}
\epsilon_{N^{\pm}_1} & =\frac{\Gamma(N^{\pm}_1\to L_{j}^{\alpha}H^{\beta})
-\Gamma(N^{\pm}_1\to\bar{L}_{j}^{\alpha}H^{\beta*})}
{\Gamma(N^{\pm}_1\to L_{j}^{\alpha}H^{\beta})
+\Gamma(N^{\pm}_1\to\bar{L}_{j}^{\alpha}H^{\beta*})}\nonumber \\
 & \approx -\frac{3}{8\pi}\sum_{k=2,3}
 \frac{{\rm Im}\big(\lambda^\dagger\lambda\big)^2_{k1}}
 {\big(\lambda^\dagger \lambda\big)_{11}}\frac{M_{1}}{M_{k}}\,,
\label{ASMN}
\end{align}
where we used the approximation $M^+_i\approx M^-_i \approx M_i$ and
we have included both the vertex contribution and the wave contribution.
We have also summed over two channels of contribution due to the $SU(2)$ doublets.
Since all of the six $N^\pm_i$ couple to $L_i H$,
the result in Eq.~(\ref{ASMN}) is twice of the result of leptogenesis models
with three heavy Majorana fields.

Since the asymmetry in the twin sector does not
communicate with the asymmetry in the SM sector,
the net $B-L$ number generated in the SM sector is given by the initially
created net lepton number $L_{\mathbf{i}}$ and thus
\begin{equation}
B-L=-L_{\mathbf{i}}=-\frac{3\,\kappa\,\epsilon\,\zeta(3)g_{N}T^{3}}{4\pi^{2}}\,,
\label{BLasy}
\end{equation}
where we define
\begin{equation}
\epsilon \equiv \epsilon_{N^{+}_1} + \epsilon_{N^{-}_1} \approx -\frac{3}{4\pi}\sum_{k=2,3}
 \frac{{\rm Im}\big(\lambda^\dagger\lambda\big)^2_{k1}}
 {\big(\lambda^\dagger \lambda\big)_{11}}\frac{M_{1}}{M_{k}}\,,
\label{ASM}
\end{equation}
and $\zeta(3)\sim1.202$, $g_{N}=2$ for the Majorana field $N^\pm_1$.
$\kappa$ is the washout factor~\cite{Buchmuller:2002rq,Buchmuller:2003gz} due to
inverse decay processes $L_{i}^{\alpha}H^{\beta},\bar{L}_{i}^{\alpha}H^{\beta*}\to N^\pm_{1}$
as well as the $2\to 2$ scattering processes.
The washout factor $\kappa$ is related to the decay parameter $K$ defined below
\bea
K \simeq \frac{\Gamma_{N^\pm_1} {\rm Br}_{N^\pm_1 \to L H}}{H(M_1)} \simeq \frac{\tilde{m}}{2 m_*}\,,
\eea
where the branching ratio is taken to be $1/2$ due to the same Yukawa couplings for both the SM and the twin leptons,
and the effective neutrino mass $\tilde{m}$ and equilibrium neutrino mass $m_*$ are defined as
\bea
\tilde{m} = \big(\lambda^\dagger\lambda\big)_{11}
\frac{v^2}{M_1}\,,
\qquad m_* = 8\pi  \frac{v^2}{M_1^2} H(M_1)\,.
\eea
The parameter $K$ characterizes the degree of whether $N^\pm_1$ is out-of-equilibrium when it starts to behave non-relativistic ($K > 1$, the strong washout regime) or not ($K < 1$, the weak washout regime).
In the weak washout limit $K \ll 1$,
the washout factor $\kappa \simeq 1$ corresponds to the most efficient case with no washout at all.
We notice that although both $\tilde{m}$ and $m_*$ contains the {\it vev} $v$,
$\kappa$ only depends on the Yukawa couplings and the lightest heavy neutrino mass $M_1$.
Since the mirror $\mathbb{Z}_2$ symmetry implies the same Yukawa couplings for both sectors,
we expect the washout factors in both the SM and the twin sector are also the same.
A detailed analysis of the washout effects
requires solving Boltzmann equations
and we leave detailed calculations for future work.

To find the relation between the
initially generated asymmetry and the current observed baryon asymmetry,
an analysis of the chemical potentials 
for particles in thermal equilibrium is required,
which allows us to compute the current value of baryon number $B_{{\rm \mathbf{f}}}$
in term of $B-L$ (since it's a conserved quantity).
For a weakly interacting plasma $\beta\mu_i\ll1$ ($\beta\equiv 1/T$),
the asymmetry in particle and antiparticle number densities is given by
\begin{equation}
n_i-\overline{n_i}\  \sim \ \frac{g_iT^{3}}{6}\times\begin{cases}
2\beta\mu_i+\mathcal{O}\big((\beta\mu_i)^{3}\big) & \qquad\qquad{\rm bosons}\,,\\
\beta\mu_i+\mathcal{O}\big((\beta\mu_i)^{3}\big) & \qquad\qquad{\rm fermions}\,,
\end{cases}\label{NDCP}
\end{equation}
where $g_i$ counts the degrees of freedom of the particle,
and $(-)\mu_i$ is the chemical potential of the (anti)particle.
Thus the asymmetries of particles being equilibrium in the thermal bath
can be expressed by their chemical potentials,
whose relation can be obtained by using the thermal equilibrium method~\cite{Harvey:1990qw,Dreiner:1992vm,Feng:2012jn},

We now investigate the thermal equilibrium in the SM sector
along with the evolution of the Universe.
We notice that while $B-L$ in the SM sector are conserved
after the decay of the heavy Majorana fields,
the baryon number $B$ is not a conserved quantity.
After the electroweak symmetry breaking,
the Higgs obtains a {\it vev},
all the SM particles in the thermal bath will then reach a new thermal equilibrium.
When top quarks drop out from the thermal bath,
all other particles in the thermal bath will reach a new balance and thus $B$ will change.
Meanwhile, sphaleron processes preserve $B-L$ but break $B$.
After the sphaleron processes become inactive,
$B$ and $L$ are separately conserved in the SM sector down to the current temperature
and correspond to the baryon and lepton number seen today which are
denoted by $B_{{\rm \mathbf{f}}}$ and $L_{{\rm \mathbf{f}}}$.

Thus to find the current baryon number in the Universe,
we focus on the temperature at which the sphaleron processes
are just going to become inactive.
This happens at a temperature of  $\sim100$~GeV which lies below
the SM top mass and we assume top quarks have already decoupled from
the thermal bath by that time.
At this temperature,
the relativistic plasma includes
the first two generations of up-type quarks and three generations of down-type quarks
($u_{iL}, u_{iR}, d_{iL},d_{iR}$),
three generations of left-handed leptons ($e_{iL}, \nu_{i}$)
and right-handed charged leptons $e_{iR}$, $i=1,2,3$,
as well as gauge bosons and the Higgs scalar $h$.
Neutral gauge bosons $Z$, photon and gluons, as well as the Higgs scalar $h$
only couple to two particles with the opposite chemical potential,
thus their chemical potentials are always zero.

The flavor mixings among quarks and neutrinos ensure
chemical potentials of the same type of fermions in different generations are equal.
Thus we drop the subscript $i$ indicating different generations.
We will use $\mu_{u_L},\mu_{u_R},\mu_{d_L},\mu_{d_R}$
to denote the chemical potentials of left-handed and right-handed up-type and down-type quarks,
$\mu_{e_L}$ and $\mu_{\nu}$ for left-handed leptons,
$\mu_{e_R}$ for right-handed charged leptons,
$\mu_{W}$ for $W^{+}$, and $\mu_{h}$ for $h$.
The following processes give rise to constraints among the chemical potentials:
The Yukawa couplings
\begin{equation}
\mathcal{L}_{{\rm Yukawa}}=g_{e_i} h \bar{e}_{iL} e_{iR}+g_{u_i} h \bar{u}_{iL} u_{iR}+g_{d_i}h \bar{d}_{iL} d_{iR}+h.c.\,,
\end{equation}
give rise to
\begin{equation}
0=\mu_{h}=\mu_{u_L}-\mu_{u_R}=\mu_{d_L}-\mu_{d_R}=\mu_{e_L}-\mu_{e_R}\,.
\label{h=0}
\end{equation}
Thus, the chemical potentials of left-handed and right-handed fermions are equal.
The gauge interactions involving $W$ bosons
($\mathcal{L}\sim W_{\mu}\bar{f}\gamma^{\mu}g+h.c.$
where $f,g$ are fermions belong to the same $SU(2)$ doublet) provide us the following relations,
\begin{align}
\mu_{W} & =\mu_{u_L}-\mu_{d_L}\qquad(W^{+}\leftrightarrow u_{L}+\bar{d}_{L})\,,\label{Y1} \\
\mu_{W} & =\mu_{\nu}-\mu_{e_L}\qquad(W^{+}\leftrightarrow \nu_{i}+\bar{e}_{iL})\,. \label{Y2}
\end{align}
The sphaleron processes give us one additional equation,
\begin{equation}
\mu_{u_L}+2\mu_{d_L}+\mu_{\nu}=0\,.\label{SPHEQ'}
\end{equation}
The neutrality of the Universe requires the total electrical charge to be zero,
with no top quark in the thermal bath
\begin{align}
4(\mu_{u_L}+\mu_{u_R}) + 6\mu_{W}- 3(\mu_{d_L}+\mu_{d_R} +\mu_{e_L}+\mu_{e_R})=0\,.
\end{align}
Solving these equations we are able to express
all the chemical potentials by the chemical potential of one particle
and we obtain the baryon and lepton number $B$ and $L$ for the temperature regime we focused on
\begin{align}
B\, & \sim\, 2(\mu_{u_L}+\mu_{u_R})+3 (\mu_{d_L}+\mu_{d_R})=-\frac{90}{19}\mu_{e}\,, \label{BC} \\
L\, & \sim\, 3\times(\mu_{e_L}+\mu_{e_R}+\mu_{\nu})=\frac{201}{19}\mu_{e}\,, \label{LC}
\end{align}
where we have dropped the same overall factor in Eq.~\ref{NDCP},
since we are only interested in the ratio of $B/(B-L)$.
Once sphaleron processes become inactive,
the $B$ and $L$ would be separately conserved,
and the ratio of $B/(B-L)$ would then freeze and show the current baryon asymmetry.
Thus, we obtain
\begin{equation}
\frac{B_{{\rm \mathbf{f}}}}{B-L}=\,\frac{30}{97}\,\approx\, 0.31\,.\label{B/B-L}
\end{equation}

The above analysis and the result of Eq.~(\ref{B/B-L})
also apply to the case of two Higgs doublet model,
as discussed in Section~\ref{sec:T2H}.
The additional Higgs gets its {\it vev} at a much higher energy than the electroweak scale
and thus has very little effect to the SM sector.
The charged Higgs from the additional Higgs doublet is heavy
and has already decoupled from the thermal bath long before
and thus also has no effect to the above analysis.

As for the twin sector of the mirror twin Higgs model,
although the twin {\it vev} is about 5.5 times of the {\it vev} in the SM sector,
focusing on the temperature regime that the twin top and twin sphaleron processes decoupled from the twin sector,
one arrives the same result,
\begin{equation}
\frac{B_{{\rm \mathbf{f}}}^\prime}{B^\prime-L^\prime}
=\,\frac{30}{97}\,\approx\, 0.31\,.\label{TB/TB-L}
\end{equation}
Though in the twin sector, the twin baryon number would freeze at a higher temperature compare to the SM sector.
For the twin two Higgs doublet model,
the second twin Higgs gets its {\it vev} at a much higher energy than the twin electroweak scale
and has almost no effects at low energies.
Thus the above analysis also applies to the twin sector from the twin two Higgs doublet model.

Now by using Eqs.~(\ref{ASM}) and (\ref{BLasy}),
one can further link the current
baryon number to the initially created net lepton number $L_{\mathbf{i}}$ as
\begin{equation}
\frac{B_{{\rm \mathbf{f}}}}{s}=\frac{30}{97}\frac{B-L}{s}=-\frac{30}{97}\frac{135\,\zeta(3)}{4\pi^{4}}\frac{\kappa\,\epsilon}{g_{s}}\,,
\label{BFSM}
\end{equation}
where the entropy density $s=2\pi^{2}g_{s}T^{3}/45$ and $g_{s}\approx100$
is the entropy degrees of freedom at $T\sim100$~GeV when the sphaleron
processes become inactive.
Using the current astrophysical constraint
given in Eq. (\ref{RBs}) we estimate $|\epsilon|\sim10^{-6}$.
As seen in Eq.~(\ref{ASM}), the asymmetry generated by the twin cogenesis mechanism
with two copies of three right-handed neutrinos, is doubled
comparing to leptogenesis with three heavy right-handed neutrinos.
Together with the neutrino masses given by Eq.~(\ref{MTHnM}),
a large parameter space of $\{M_{i}^\pm, \lambda_{ij}\}$ is allowed to achieve
the right amount of asymmetry as well as the desired neutrino masses less than
$\mathcal{O}({\rm eV})$.
For example, $M_1\sim 10^{14}$~GeV, $M_1/M_k \sim 10^{-2}$
and $\lambda_{ij}\sim \mathcal{O}(10^{-2})$ lead to consistency with
the current baryon asymmetry and the tiny neutrino masses.

The same leptogenesis mechanism applies also to the twin sector.
Due to the out-of-equilibrium decay of $N^\pm_{i}$,
a net twin lepton number $L_{\mathbf{i}}^{\prime}$
was generated and then transferred to the
twin baryon sector through the twin sphaleron processes.
The processes generating the twin lepton number
are similar to
the processes in the SM sector discussed earlier.
We notice that although there is an additional minus sign for the Yukawa coupling
of $-\lambda_{ij} \bar{N}^-_{i}L_{j}^{\prime}H^{\prime}$ term from Eq.~(\ref{AsyGenII}),
this coupling always comes in pairs
in the calculation involving $N^-_{i}$
and thus the additional minus signs always cancel out.
Thus the result of Eq.~\ref{ASM} also holds for the twin sector,
generating an excess of $L_{i}^{\prime \alpha}H^{\prime \beta}$
over $\bar{L}_{i}^{\prime \alpha}H^{\prime \beta*}$.

Since the complex coupling $\lambda_{ij}$
are identical in the SM and the twin sector,
and under the assumption that the washout effects in the twin sector are the same as the SM sector,
the initially created lepton number and twin lepton number have a similar amount.
Thus a similar amount of net $B^\prime-L^\prime$ number is generated also in the twin sector,
\begin{equation}
B^\prime-L^\prime=-L_{\mathbf{i}}^\prime=-\frac{3\,\kappa\,\epsilon\,\zeta(3)g_{N}T^{3}}{4\pi^{2}}\,,
\label{TBLasy}
\end{equation}
where $L_{\mathbf{i}}^\prime$ is the initially created net twin lepton number.
When the temperature drops down,
after the twin tops decouple from the thermal bath
and twin sphaleron processes also become inactive,
the twin baryon number freezes to $B_{{\rm \mathbf{f}}}^{\prime}$
given by Eq.~(\ref{TB/TB-L}).
Using Eqs.~(\ref{BFSM}) and (\ref{TBLasy}), one finds
in the twin sector the
final twin baryon number is approximately equal to the SM baryon number, i.e.,
\begin{equation}
B_{{\rm \mathbf{f}}}^{\prime}\approx B_{{\rm \mathbf{f}}}\,.
\end{equation}
As mentioned in Section~\ref{sec:MTH},
twin fermions which are heavier than twin electrons,
can annihilate to twin electron-positron pairs through twin electroweak interactions.
In this way the symmetric component of twin matter fermions are removed.
Thus we are left with the lightest asymmetric twin baryons as the dark matter candidates.
Therefore, if the masses of the lightest twin baryons are compatible with the mass of a proton,
the puzzle of cosmic coincidence Eq.~(\ref{RDMB}) can be naturally explained.

In the full Higgs sector, the approximate $SU(4)$ global symmetry
is spontaneously broken when the twin Higgs obtain a {\it vev} $f$,
and one of the pseudo-Nambu-Goldstone bosons is identified to be the SM
Higgs, which acquires {\it vev} at the measured value $v$.
The other six degrees of freedom become the longitudinal part
of the $W^{\pm}, Z, W^{\prime \pm}, Z^{\prime}$ gauge bosons.
This pseudo-Nambu-Goldstone boson should be consistent with the Higgs observed at the LHC,
which sets a constraint on the twin Higgs {\it vev} $f/v\geq3$~\cite{Khachatryan:2016vau}.
With the whole symmetric component of twin particles annihilated efficiently,
the (asymmetric) twin baryons are the dark matter candidates.
The mass of twin baryons to be 5.5~GeV solves the cosmic coincidence puzzle
$(\Omega_{{\rm DM}}h^{2})/(\Omega_{{\rm B}}h^{2})\approx5.5$ automatically.
Thus the twin cogenesis requires the twin QCD confinement scale to be
$\Lambda^\prime_{\rm QCD} \approx 5.5 \,\Lambda_{\rm QCD}$.

\section{Twin cogenesis for the fraternal twin Higgs model}\label{sec:CogFTH}

The twin cogenesis can also be realized in a fraternal twin Higgs setup~\cite{Craig:2015pha}.
The matter content in the fraternal twin sector includes
(left-handed and right-handed) twin top and twin bottom quarks
($Q^{\prime},t_{R}^{\prime},b_{R}^{\prime}$),
twin tau and left-handed twin tau neutrino ($L_\tau^{\prime},\tau_{R}^{\prime}$).
In this case we introduce three heavy right-handed neutrinos $N_i$
with complex Yukawa coupling constants $\lambda_{ij}$ ($i,j=1,2,3$),
which are responsible for the tiny neutrino masses in the SM.
For this case the Yukawa interactions involving
the three heavy Majorana fermions $N_i$ are given by Eq.~(\ref{AsyGenFTH}).
The neutrino masses are generated from the seesaw mechanism,
while the twin tau neutrino mass requires to diagonalize a $4\times 4$ mass matrix.
Assuming the mass of $N_{2},N_{3}$ are much larger than the mass of $N_{1}$,
only the mass of the lightest Majorana neutrino $N_1$ plays a role.
The neutrino and twin neutrino masses are given by
\begin{equation}
m_{ij}  \simeq  \lambda_{ik} \left(M_k^{-1} \right) \lambda_{kj} v^2\,,
\qquad m_{\nu'_\tau} \simeq  |{\lambda_{13}}|^2 \frac{f^2}{M_1^2}\,.
\label{MTHnM}
\end{equation}
Due to the seesaw mass mixing, the mass of the tau neutrino is always zero.\footnote{
In addition to the three heavy right-handed Majorana neutrinos $N_i$,
one can also introduce a fourth $N^\prime$ coupling to $L_3^{\prime}H^{\prime}$,
similar to the setup as Eq.~(\ref{AsyGen2N}).
In this case there will not be a massless neutrino,
and the analysis would be also slightly different.}

The contribution to leptogenesis in the fraternal twin Higgs setup
also include the tree level decay $N_1 \to L_3^\prime H^\prime$
interference with the vertex and wave function one-loop diagrams.
In this case $N_i$ only couple to one generation of twin leptons, and
the excess of $L_{3}^{\prime \alpha}H^{\prime \beta}$
over $\bar{L}_{3}^{\prime \alpha}H^{\prime \beta*}$ is given by~\cite{Feng:2013zda}
\begin{align}
\epsilon_{\,\rm FTH} & =\frac{\Gamma(N_{1}\to L_{3}^{\prime\alpha}H^{\prime\beta})
-\Gamma(N_{1}\to\bar{L}_{3}^{\prime\alpha}H^{\prime\beta*})}
{\Gamma(N_{1}\to L_{3}^{\prime\alpha}H^{\prime\beta})
+\Gamma(N_{1}\to\bar{L}_{3}^{\prime\alpha}H^{\prime\beta*})}\nonumber \\
 & \approx-\frac{3}{16\pi}\sum_{k=2,3}
 \frac{{\rm Im}(\lambda_{13}^{2}\lambda_{k3}^{*2})}{|\lambda_{13}|^{2}}\frac{M_{1}}{M_{k}}\,.\label{FTHASM}
\end{align}
Thus the $B^\prime-L^\prime$ asymmetry generated in the twin sector of the fraternal twin Higgs setup
is given by
\begin{equation}
(B^\prime-L^\prime)_{\rm FTH} = -L_{\mathbf{i}\,{\rm FTH}}^\prime=
-\frac{3\,\kappa_{\,\rm FTH}\,\epsilon_{\,\rm FTH}\,\zeta(3)g_{N}T^{3}}{4\pi^{2}}\,,
\label{FTHBLasy}
\end{equation}
where $L_{\mathbf{i}\,{\rm FTH}}^\prime$ is the initially created net twin lepton number for the fraternal Higgs model
and $\kappa_{\,\rm FTH}$ is the washout factor in the fraternal twin sector.

Now we focus on the thermal history of the twin sector for the
fraternal twin Higgs case. 
After the twin electroweak phase transition takes place,
the twin top gets its mass and gradually becomes non-relativistic
and starts to decouple from the thermal bath.
At the same time the twin sphaleron processes also become inactive,
after which the net baryon number excess will freeze inside the twin sector.
Since there is only one generation of baryon,
twin tops can only decay to twin bottoms without changing the total twin baryon number.
Thus for the case of only one generation fermions, the net baryon number computed
when all the twin sector particles are in thermal equilibrium,
will not change all the way down to the current temperature.
In particular, the ratio $B^\prime/(B^\prime-L^\prime)$ will not change.
This is very different with the case of three generations of fermions discussed in Section~\ref{Sec:CogMTH}.
In the case of three generations,
when the top decouples from the thermal bath,
the sphaleron processes are still active for the first two generations of quarks
and top quarks can also decay to quarks in the first two generations through flavor mixings,
thus the neutrality condition requires the total electric charge of
all charged particles without top quark equal to zero.

Now let's compute the total baryon number excess in the twin sector.
When all twin sector particles are in thermal equilibrium,
particles contributing to the chemical potential equations are:
the left-handed twin quarks and twin leptons ($t_L^\prime, b_L^\prime, \tau_L^\prime, \nu_{\tau}^\prime$),
the right-handed twin quarks and twin taus ($t_R^\prime, b_R^\prime, \tau_R^\prime$),
and twin $W^{\prime \pm}$ bosons.
The chemical potential equations are
\begin{gather}
 \mu_{t^\prime_L}-\mu_{b^\prime_L}=\mu_{\nu_{\tau}^\prime}-\mu_{\tau_L^\prime} = \mu_{W^\prime}\,,\\
\mu_{t_L^\prime}-\mu_{t_R^\prime}
=\mu_{b_L^\prime}-\mu_{b_R^\prime}=\mu_{\tau_L^\prime}-\mu_{\tau_R^\prime}=  0\,,\\
\mu_{t_L^\prime}+2\mu_{d_L^\prime}+\mu_{\nu_\tau^\prime}=  0\,,\\
6\mu_{W^\prime}+2(\mu_{t^\prime_L}+\mu_{t_R^\prime})
- (\mu_{d_L^\prime}+\mu_{d_R^\prime} +\mu_{e_L^\prime}+\mu_{e_R^\prime})=  0\,.
\end{gather}
With the same overall factor, the baryon and lepton numbers can be written in terms of $\mu_{\tau_L^\prime}$
\begin{equation}
B \sim  -\frac{20}{13}\,\mu_{\tau_L^\prime}\,,\qquad
L \sim  \frac{43}{13}\,\mu_{\tau_L^\prime}\,.
\end{equation}
And thus for the one generation Fraternal twin Higgs sector
\begin{equation}
\left(\frac{B_{\mathbf{f}}^\prime}{B^\prime-L^\prime} \right)_{\rm FTH}= \frac{20}{63} \approx 0.32\,.
\label{FTHBF}
\end{equation}
Using Eqs.~(\ref{BFSM}), (\ref{FTHBLasy}) and (\ref{FTHBF}),
one arrives for the fraternal Higgs model,
\begin{equation}
\frac{B_{\mathbf{f}}}{B_{\mathbf{f}\,{\rm FTH}}^\prime} = \frac{189\,\kappa\, \epsilon}{194\,\kappa_{\rm FTH}\,\epsilon_{\,\rm FTH}}\,.
\label{B/BFTH}
\end{equation}

As discussed above,
for the fraternal twin Higgs model with one generation of twin fermions,
when temperature drops down to the current temperature,
we are left with twin bottom quarks with mass $(f/v)\,m_b$, twin taus with mass $(f/v)\,m_\tau$,
and twin tau neutrinos with mass $(f/v)\,m_{\nu_\tau}$ which will appear as dark radiation.
Also as discussed in Section~\ref{Sec:CogMTH},
a massive dark photon with mass $\sim 10$~MeV
can efficiently deplete the symmetric component of twin sector fermions to electron-positron pairs.
Thus the dark matter candidate in this case
is the twin bottom bound states $\Omega^\prime_{b^\prime b^\prime b^\prime}$
with $U(1)^\prime_{\rm em}$ charge $-1$,
or probably $\Omega^\prime_{b^\prime b^\prime b^\prime}$
bounded with a twin anti-tau forming a twin atom with total $U(1)^\prime_{\rm em}$ charge 0~\cite{Garcia:2015toa}.
Computations in \cite{Yang:2019lsg} show $m_{\Omega_{b b b}} \approx 14$~GeV.
Assuming no further dynamics above the scale of $\Lambda=4\pi f$
and no additional $\mathbb{Z}_2$ breaking terms,
the mass of the twin triple bottom baryon $\Omega^\prime_{b^\prime b^\prime b^\prime}$ would be
similar to the mass of  $\Omega_{b b b}$ around 14~GeV.
In general, the mass of $\Omega^\prime_{b^\prime b^\prime b^\prime}$ would be larger with additional physical effects.
Setting $\kappa_{\,\rm FTH}$ identical to the washout factor in the SM following the discussion in Section~\ref{Sec:CogMTH},
and using Eqs.~(\ref{RDMB}) and (\ref{B/BFTH}), the fraternal twin cogenesis requires
\begin{equation}
\epsilon_{\,\rm FTH} \sim 5.5\, \epsilon\,,
\end{equation}
which can be easily satisfied by a proper choice of the coupling constants $\lambda_{ij}$ in Eq.~(\ref{AsyGenFTH}).

\section{Phenomenology}\label{sec:Pheno}

In twin cogenesis, the SM sector and the twin sector are connected through three types of portal particles:
the Higgs boson, the dark photon, and the heavy Majorana neutrinos.
These portal interactions will leave traces in different experiments:
the LHC, low energy experiments, dark matter searches, and neutrino experiments.
Among all these,
direct detections of 5.5~GeV dark matter candidates
will provide smoking-gun signatures for the twin cogenesis mechanism.

\subsection{Higgs signal strength and Higgs exotic decays}


The Higgs boson provides the connection between visible and twin sector
since it couples to both the SM particles and the twin particles.
Due to its pseudo-Nambu-Goldstone boson nature, there is a small misalignment between the SM Higgs boson and the pseudo-Nambu-Goldstone boson, parameterized by the parameter $\theta = v/f$ with $\cos\theta = \sqrt{1- v_{\rm ew}^2/f^2}$ and the electroweak {\it vev} $v_{\rm ew} \equiv f \sin \theta$.
The tree-level couplings of the Higgs boson to the fermions and bosons in the SM (twin) sector are altered by a factor $\cos\theta$ ($\sin\theta$) relative to the SM ones.
In particular, the coupling of the physical Higgs boson to vector bosons is given by
\begin{align}
\mathcal{L}_{hVV}&= \left[\frac{g^2 v_{\rm ew}^2}{2}W_{\mu}^+W^{\mu,-}+\frac{(g^2+g'^2) v_{\rm ew}^2}{4}Z_{\mu}Z^{\mu}\right] \frac{h}{v_{\rm ew}} \cos\theta \nn\\
&\quad -\left[\frac{g^2 f^2 \cos^2\theta }{2}W_{\mu}^{\prime +}W^{\prime\mu,-}+\frac{(g^2+g'^2) f^2 \cos^2\theta}{4}Z'_{\mu}Z^{\prime\mu}\right] \frac{h}{f \cos\theta} \sin\theta\,,
\end{align}
and the Yukawa couplings read
\begin{equation}
\mathcal{L}_{hff} = {\rm i} y_f \bar{f} f h \cos\theta - {\rm i} y_f \bar{f'} f' h \sin\theta + h.c.\,.
\end{equation}
Thus all the SM Higgs production and decay cross sections are modified by a common factor
\begin{equation}
	\sigma_{pp \to h} = \cos^2\theta \sigma^{\rm SM}_{pp \to h}\,, \qquad
\Gamma_{h \to {\rm SM}_i} = \Gamma^{\rm SM}_{h \to {\rm SM}_i}\cos^2\theta\,.
\end{equation}

\begin{figure}[!h]
\center{\includegraphics[width=0.5\textwidth]
{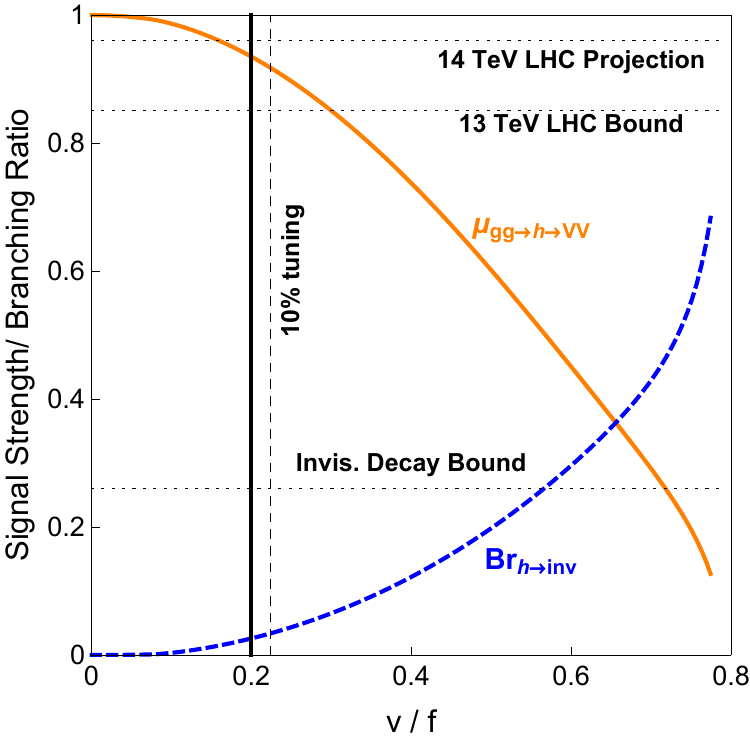}}
\caption{The signal strength in gluon fusion production and subsequent decays to gauge boson pairs, and the Higgs invisible branching ratio as function of $v/f$. The bound on the invisible decay branching ratio is
${\rm Br}_{\rm inv} < 0.26$ based on the combined ATLAS results. }
\label{fig:higgsfit}
\end{figure}

There are also exotic Higgs decay channels which contribute to the invisible decay signatures and/or displaced vertex signatures.
Given the twin spectrum, the invisible decay rate into twin fermions is
\begin{equation}
\Gamma_{\rm inv} = \sum_{f' = e', \cdots, b'}\Gamma (h\to  \bar{f'} f') + \Gamma (h\to g'g')\,,
\end{equation}
here we note that the off-shell decay channel $h \to V'V' \to 4f$ is negligibly small due to the suppressions from factor $v^2/f^2$ and off-shell kinematics.
The partial decay widths of the Higgs boson into twin fermions are given by
\begin{align}
\Gamma (h\to \bar{f'}f')=N_C\frac{y_f^2 \sin^2\theta}{8\pi}m_h \left[1-\frac{4m_{f}^{\prime 2}}{m_h^2}\right]^{3/2}\,,
\label{eq:decayfermion}
\end{align}
where $N_C$ is the color factor $N_C=3(1)$ for mirror quarks (leptons).
The decay width of the Higgs boson into twin gluons is written as, via twin quark loops,
\begin{align}
\Gamma (h\to g'g')= \sum_{f'}\frac{y_{f'}^{2} \sin^2\theta  \alpha_s^{\prime 2} m_h^3}{128\pi^3  m_{f}^{\prime 2}}\left|F_{1/2}\left(\frac{4  m_{f}^{\prime 2}}{m_h^2}\right)\right|^2\,,
\end{align}
where the loop function is defined as
\begin{align}
F_{1/2}(\tau)\equiv -2\tau[1+(1-\tau)f(\tau)]\,,
\end{align}
along with
\begin{equation}
  f(\tau)=\begin{cases}
    [\sin^{-1}\sqrt{1/\tau}]^2, & \text{$\tau\geq 1$}\,,\\
    -\frac{1}{4}\big[\log\frac{1+\sqrt{1+\tau}}{1-\sqrt{1-\tau}}-{\rm i}\pi\big]^2, & \text{$\tau<1$}\,.
  \end{cases}
\end{equation}
These exotic decay channels dominantly contribute to the Higgs invisible decay.

We calculate various Higgs signal strengths $\mu_{pp \to h \to ii} = \sigma_{pp \to h } {\rm Br}_{h \to ii}/ (\sigma_{\rm SM} {\rm Br}_{\rm SM})$ and the Higgs invisible decay width.
Among all the Higgs production and decay channels, the gluon fusion production and gauge boson pair decay channel is the most precise one, and thus we will use this channel to put limit on the parameter $v/f$.
The latest ATLAS result on this channel~\cite{ATLAS:2020rej} sets bound on the signal strength $\mu_{gg \to h \to ZZ}  = 0.96 \pm 0.1 (\textrm{stat.}) \pm 0.03 (\textrm{sys.}) \pm 0.03 (\textrm{th.})$ from the data at the 139 fb$^{-1}$ luminosity.  A rough bound on $v/f$ smaller than 0.3 at the 95\% confidence level could be obtained.
The high luminosity LHC will improve sensitivity of signal strengths to around 4\%~\cite{ATLAS:2018jlh} assuming current uncertainty with 3 ab$^{-1}$ luminosity.
%
Finally we investigate the Higgs invisible decay, which is another distinct signature in the twin Higgs model. According to current LHC searches~\cite{Aaboud:2019rtt}, the upper limit on the invisible decay branching ratio is set to be ${\rm Br}_{\rm inv} < 0.26$.
The signal strength in gluon fusion production and subsequent decays to gauge boson pairs, and the Higgs invisible branching ratio as function of $v/f$ are shown in Fig.~\ref{fig:higgsfit}.
We expect future experiments such as MATHUSLA and FASER~\cite{Curtin:2018mvb, Feng:2017uoz, Alimena:2019zri} can probe the Higgs exotic decay channels through the displaced vertices signatures.

\subsection{Dark photon mixing}

The mixings between the SM sector and the twin sector are given by~\cite{Feldman:2007wj}
\begin{equation}
\mathcal{L_{{\rm mix}}}=-\frac{\epsilon}{2}F_{\mu\nu}F^{\prime\mu\nu}-\frac{1}{2}(m B_{\mu}+m^{\prime}B_{\mu}^{\prime}+\partial_{\mu}\sigma)^{2}\,,
\end{equation}
where $F_{\mu\nu}$ and $F_{\mu\nu}^{\prime}$ are the $U(1)_{Y}$
and $U(1)_{Y}^{\prime}$ field strengths with corresponding gauge
bosons $B_{\mu}$ and $B_{\mu}^{\prime}$ respectively. The kinetic
mixing and mass mixing allow the dark photon to couple to the SM fermions.
In the gauge eigenbasis $G^{T}=(B,B^{\prime},A_{3},A_{3}^{\prime})$
of neutral gauge bosons, the mixing matrices can be written as
\begin{equation}
\mathcal{K}=\left(\begin{array}{cccc}
1 & \epsilon & 0 & 0\\
\epsilon & 1 & 0 & 0\\
0 & 0 & 1 & 0\\
0 & 0 & 0 & 1
\end{array}\right),
\quad
M_{{\rm St}}^{2}=\left(\begin{array}{cccc}
m^{2}+\frac{1}{4}g_{Y}^{2}v^{2} & mm^{\prime} & -\frac{1}{4}g_{2}g_{Y}v^{2} & 0\\
mm^{\prime} & m^{\prime2}+\frac{1}{4}g_{Y}^{2}f^{2} & 0 & -\frac{1}{4}g_{2}g_{Y}f^{2}\\
-\frac{1}{4}g_{2}g_{Y}v^{2} & 0 & \frac{1}{4}g_{2}^{2}v^{2} & 0\\
0 & -\frac{1}{4}g_{2}g_{Y}f^{2} & 0 & \frac{1}{4}g_{2}^{2}f^{2}
\end{array}\right),
\end{equation}
where $\mathcal{K}$ and $M_{{\rm St}}^{2}$ are the kinetic mixing
and Stueckelberg mass mixing matrices respectively. A simultaneous
diagonalization of the above two matrices allows one to work in the
mass eigenbasis $E^{T}=(A,A^{\prime},Z,Z^{\prime})$, where $A,A^{\prime}$
are the photon and dark photon fields respectively. To simplify the
results, firstly we assume $m=0$ such that there is no mixing from
the Stueckelberg sector. Since the dark photon mass $\sim m^{\prime}\sim 10$~MeV
is much smaller than the Higgs or twin Higgs induced mass terms,
the Stueckelberg mass matrix $M_{{\rm St}}^{2}$ can be then approximately
reduces to the two following matrices
\begin{equation}
M_{{\rm Twin}}^{2}=\left(\begin{array}{cc}
m^{\prime2}+\frac{1}{4}g_{Y}^{2}f^{2} & -\frac{1}{4}g_{2}g_{Y}f^{2}\\
-\frac{1}{4}g_{2}g_{Y}f^{2} & \frac{1}{4}g_{2}^{2}f^{2}
\end{array}\right),\quad
M_{{\rm StSM}}^{2}=\left(\begin{array}{ccc}
\frac{1}{4}g_{Y}^{2}v^{2} & 0 & -\frac{1}{4}g_{2}g_{Y}v^{2}\\
0 & m^{\prime2} & 0\\
-\frac{1}{4}g_{2}g_{Y}v^{2} & 0 & \frac{1}{4}g_{2}^{2}v^{2}
\end{array}\right).
\end{equation}
From the diagonalization of the twin sector gauge boson mass matrix
$M_{{\rm Twin}}^{2}$, one recovers a massless dark photon $\gamma^{\prime}$
and a massive $Z^{\prime}$ with mass $m_{Z^\prime} =f\sqrt{g_{2}^{2}+g_{Y}^{2}}/2$
for the case $m^{\prime}=0$, with gauge eigenbasis $( B^\prime, A_3^\prime)$.
With the presence of the $m^{\prime2}$
term, $M_{{\rm Twin}}^{2}$ introduce additional mixing between $\gamma^{\prime}$
and $Z^{\prime}$ in the twin sector, which allows the dark photon
to couple to twin neutrinos. Since $m^{\prime}\ll g_{2}f,g_{Y}f$,
one can approximate the new mass eigenbasis of $M_{{\rm Twin}}^{2}$
in terms of the original mass eigenbasis $(A^{\prime},Z^{\prime})$ for
the case $m^{\prime}=0$.
More explicitly, one has
\begin{equation}
A_{\mu}^{\prime(m)}\approx
A_{\mu}^{\prime}
+\frac{m^{\prime2}}{f^{2}}\frac{2 \sin2\theta_W}{g_{2}^{2}+g_{Y}^{2}}\, Z_{\mu}^{\prime}\,,
\end{equation}
where $\theta_W$ is the Weinberg angle, and the massive dark photon
now carries a small fraction of $Z^{\prime}$.
Below we will drop the superscript $(m)$ in $A_{\mu}^{\prime(m)}$.
Given the coupling
of twin $Z^{\prime}$ to twin neutrinos
\begin{equation}
\mathcal{L}_{Z^{\prime}\bar{\nu}^{\prime}\nu^{\prime}}
\sim\frac{g_{2}}{2\cos\theta_W}Z_{\mu}^{\prime}\bar{\nu}^{\prime}_L\gamma^{\mu}\nu^{\prime}_L\,,
\end{equation}
the coupling of the massive dark photon to twin neutrinos can be
written as
\begin{equation}
\mathcal{L}_{\gamma^{\prime}\bar{\nu}^{\prime}\nu^{\prime}}
\sim \frac{m^{\prime2}}{f^{2}}\frac{\sin2\theta_W}{\sqrt{g_{2}^{2}+g_{Y}^{2}}}
A_{\mu}^{\prime}\bar{\nu}^{\prime}_L\gamma^{\mu}\nu^{\prime}_L\,.\label{DPhTn}
\end{equation}
On the other hand,
the simultaneous diagonalization of the kinetic mixing matrix $\mathcal{K}$ and $M_{{\rm StSM}}^{2}$
with gauge eigenbasis $(B, B^\prime, A_3)$,
gives rise to mixings between the dark photon with the SM photon and $Z$ boson,
which was well-studied in~\cite{Feldman:2007wj}.

In this work, we are interested in the coupling of the dark photon
to SM neutrinos $\gamma^{\prime}\bar{\nu}\nu$, and the coupling
of the SM $Z$ boson to twin neutrinos $Z\bar{\nu}^{\prime}\nu^{\prime}$.
To find these couplings, one needs the fraction of $Z$ in $B^{\prime}$
and the fraction of $A^{\prime}$ in $A^{3},$which are given by
\begin{align}
B^{\prime} & =\left[\cos\psi\sin\theta_W\sin\phi+\cos\phi\sin\psi
+\textstyle{\frac{\epsilon}{\sqrt{1-\epsilon^{2}}}} (\cos\phi\cos\psi\sin\theta_W-\sin\phi\sin\psi)\right]Z+\cdots\,,\\
B\  & = \textstyle{\frac{1}{\sqrt{1-\epsilon^{2}}}} (\cos \psi \sin \phi + \cos \phi \sin \theta_W \sin \psi )\,A^{\prime} + \cdots\,,\label{ApinB}\\
A^{3} & =-\cos\theta_W\sin\psi\ A^{\prime}+\cdots\,,\label{ApinA3}
\end{align}
where $\cdots$ denotes other components of $B, B^{\prime}, A^{3}$ in terms of
either the photon, the dark photon, or $Z$ gauge boson in the mass eigenbasis,
which would be irrelevant for the discussion here.
$\theta_W$ is again the Weinberg angle, and angles $\phi,\psi$ are given by
\begin{equation}
\tan\phi =\frac{-\epsilon}{\sqrt{1-\epsilon^{2}}}\,,
\qquad
\tan2\psi \approx 2 \epsilon \sqrt{1-\epsilon^{2}} \sin\theta_W \,.
\end{equation}
Thus the $\gamma^{\prime}\bar{\nu}\nu$ and $Z\bar{\nu}^{\prime}\nu^{\prime}$
couplings read
\begin{align}
\mathcal{L}_{\gamma^{\prime}\bar{\nu}\nu} & \sim -\frac{1}{2}
\left(g_Y \tan \phi + \frac{g_2}{\cos \theta_W} \sin \psi\right)
A_{\mu}^{\prime}\bar{\nu}_L\gamma^{\mu}\nu_L\,, \label{DPhn}\\
\mathcal{L}_{Z\bar{\nu}^{\prime}\nu^{\prime}} & \sim
\frac{\sin2\theta_W}{\sqrt{g_{2}^{2}+g_{Y}^{2}}}
\sin\psi \,\frac{m^{\prime2}}{f^{2}}\,
Z_{\mu}\bar{\nu}^{\prime}_L\gamma^{\mu}\nu^{\prime}_L\,, \label{ZTn}
\end{align}
where we have used Eq. (\ref{DPhTn}) and we also dropped terms of order $\mathcal{O}(\epsilon^{2})$.
In Eq. (\ref{ZTn}) we find the SM $Z$ boson couple to twin neutrinos with
a double suppression $\mathcal{O}(m^{\prime2}/f^{2})\mathcal{O}(\epsilon)\sim\mathcal{O}(10^{-20})$,
thus we expect the invisible decay of $Z\to\bar{\nu}^{\prime}\nu^{\prime}$ shows very tiny effect.
The above mixings also allow the twin neutrinos annihilating
to SM electron-positron pairs through the resonance production of
the dark photon. However, both of the dark photon to twin neutrino coupling
and the dark photon to SM neutrino coupling are induced by tiny mixings,
which make $\bar{\nu}^{\prime}\nu^{\prime}\to\bar{e}e$ process order
of $\mathcal{O}(10^{-20})$ weaker than normal electroweak interactions.
Thus the twin neutrinos can be depleted only via the two-step process,
i.e., twin neutrinos annihilate to twin electron-positron pairs through
twin electroweak interactions and twin electron-positron pairs subsequently
annihilate to electron-positron pairs through the dark photon portal.
We also notice that with no mass mixing from the Stueckelberg sector,
the massless SM photon does not couple to hidden sector fermions.
Thus the twin sector fermions carry exactly zero electric charge.

The experimental limits set on the dark photon mixing mostly come from
the dark photon to electron-positron coupling,
which is given by
\begin{equation}
\mathcal{L}_{\gamma^{\prime}\bar{e}e}
\sim
A^\prime_\mu \bar{e} \gamma^\mu ( v^\prime_e - \gamma_5 a^\prime_e ) e\,,
\end{equation}
where the vector and axial couplings can be computed using Eqs.~(\ref{ApinB}) and (\ref{ApinA3})
\begin{align}
v^\prime_e & =\frac{g_2}{4} \cos\theta_W \sin \psi  - \frac{3 g_Y}{4} \sin\theta_W \sin \psi
- \frac{3 g_Y }{4} \tan \phi\,, \\
a^\prime_e & = \frac{g_2}{4} \cos\theta_W  \sin \psi + \frac{g_Y}{4} \sin\theta_W \sin \psi
+ \frac{g_Y}{4} \tan \phi\,.
\end{align}
Thus the $\gamma^{\prime}\bar{e}e$ coupling is about a factor of
$\mathcal{O}(\epsilon)$ weaker than the electromagnetic interaction.
\cite{Fabbrichesi:2020wbt} discussed comprehensive bounds on the dark photon couplings.
For the $\mathcal{O}(10)$~MeV dark photon, the limit is $\epsilon \lesssim 10^{-9}$.
The $Z^\prime \bar{f}f$ coupling (here $f$ denotes all SM fermions),
however, is further induced by the diagonalization of $M_{{\rm Twin}}^{2}$ matrix,
which would be again a double suppression
$\mathcal{O}(m^{\prime2}/f^{2})\mathcal{O}(\epsilon)\sim\mathcal{O}(10^{-20})$
compare to electroweak interactions.
Given the mass of $Z^\prime$ around 500~GeV,
these interactions do not impose further constraint on the model.

\subsection{Dark matter detections}

As mentioned in previous sections, the dark matter would be the lightest dark baryons,
i.e., twin protons and twin neutrons with masses around 5.5 GeV in the mirror twin Higgs cogenesis.
For the fraternal twin Higgs cogenesis,
the dark matter candidate is the twin triple bottom baryon $\Omega^\prime_{b^\prime b^\prime b^\prime}$
with mass $\gtrsim 14$~GeV.
Here we will focus on the 5.5~GeV twin baryons as the dark matter candidates.
Their portal couplings to the SM are through the Higgs and the dark photon.
On one hand, since it was produced in early universe asymmetrically,
the indirect detection signature would be sub-dominant.
On the other hand,
direct detection experiments such as PandaX, XENON1T, CDMS-lite, etc,
may have the possibility to detect such dark matter in the future.
Here we calculate the direct detection rate for the Higgs and dark photon portals.

The direct detection rate from the Higgs portal is estimated to be
\bea
	\sigma_{\rm SI} = \frac{1}{16\pi}\left(\frac{4 m_\chi m_N}{m_\chi + m_N}\right)^2 \frac{m_N^2  f_N^2}{v^2} \frac{1}{m_h^4} \frac{m_\chi^2 f_\chi^2}{f^2} \frac{v^2}{f^2} \,,
\eea
where the dark matter $\chi$ is identified as the twin nucleon in the mirror twin Higgs models.
The effective couplings to the nucleon $F_N$ and the twin nucleon $F_\chi$
are related to the nucleon scalar current matrix elements as follows
\bea
	\langle N| m_q \overline{\psi_q}\psi_q| N \rangle = f_q^N m_N\,, \qquad
	\langle N'| m_q \overline{\psi_{q'}}\psi_{q'}| N' \rangle = f_{q'}^{N'} m_N\,.
\eea
Thus the nucleon effective coupling is calculated to be
\bea
	f_N = \frac{2}{9} + \frac{7}{9} \sum_{q= u,d,s} f_{Tq}\,,
\eea
and similarly the twin nucleon effective coupling $f_\chi$ can be estimated accordingly.
For the 5.5~GeV twin baryon dark matter,
the nucleon cross-section is estimated to be around $10^{-48}$ cm$^2$,
much lower than the latest bounds from the XENON1T result~\cite{Aprile:2018dbl}.

Since the mass of the dark photon is around 10~MeV,
dark matter and nuclear scattering mediated by the dark photon can be significant.
The spin-independent cross-section due to the light dark photon is estimated to be
\bea
	\sigma_{\rm SI} = \frac{1}{16\pi}\left(\frac{4 m_\chi m_N}{m_\chi + m_N}\right)^2 \frac{G^2_\chi G^2_N}{(2 m_N E_R + m_{\gamma'}^2)^2}\,,
\eea
where the effective couplings to the nucleon are $G_p = 2 g_{vu} + g_{vd}$ and $G_n = g_{vu} + 2g_{vd}$, and similarly the $G_\chi$ for the effective coupling to the twin nucleon.
We refer to \cite{Belanger:2020gnr, Hambye:2018dpi} for more detailed analysis.
Using the upper bound on the kinetic mixing parameter $\epsilon \sim 10^{-9}$
from $\mathcal{O}(10)$~MeV dark photon searches~\cite{Fabbrichesi:2020wbt},
an estimation shows $\sigma_{\rm SI} \sim 10^{-47}~{\rm cm}^2$,
which is just on the edge of the sensitivity of the XENONnT experiment.
Thus improved experiments in the future in the low dark matter mass region
with better sensitivities have the possibility
to test twin cogenesis mechanism with 5.5~GeV twin baryons as the dark matter.

\subsection{Twin neutrino mixing and decay}

Given the neutrino Yukawa interactions in Eqs.~(\ref{AsyGen2N}) and (\ref{v2HDMYuk}), the right-handed neutrinos $N^\pm$ induce the mixing between the SM neutrinos and the twin neutrinos.
Let us first work out the mixing angles between the SM and the twin neutrinos in the mirror twin Higgs model.
According to the Eqs.~(\ref{AsyGen2N}) and (\ref{v2HDMYuk}),
the mass matrices in the $(\nu_L, \nu'_L, N^c, N^{\prime c})$ basis is
\bea
	{\cal M}_{\nu} = \left(
	\begin{array}{cccc}
		0 & 0 & \lambda u & 0 \\
		0 & 0 &  0 & \lambda f \\
		\lambda^\dagger u & 0 & M & m \\
		0 & \lambda^\dagger f & m^T & M \\
	\end{array}
	\right)\,,
\eea
where each element contains $3 \times 3$ matrix for three generations.
To diagonalize the mass matrix, we perform several step rotations: first we rotate the ${\cal M}_{\nu} $ to have the lower $2\times 2$ matrix block diagonal,
then do further rotation to block diagonal the off-diagonal $\lambda$ terms, finally a rotation to diagonal the SM neutrino and the twin neutrino masses.
Since the details are not so relevant here, we just show the estimation of the final expression for the mixing angle between the SM neutrinos and the twin neutrinos
\bea
	\theta_{\nu\nu'} \simeq \lambda_{ik} \lambda^*_{kj}\frac{u f_2}{M_k^2} \left[ 1+ {\cal O}\left(\frac{m^2_{\rm d}}{M^2}\right) \right] + \cdots\,.
\eea
According to the mass insertion technique, there are two mass insertions between the SM neutrinos and the twin neutrinos, and thus the mixing angles are quite suppressed.
Since we assume $m_{\nu'} \ge m_{e} $, there is only one decay channel for the twin neutrino decay channel $\nu' \to \nu e^+ e^-$ via the off-shell $Z$ exchange.
Because of the suppression for the mixing angle, unlike the case in~\cite{Chacko:2016hvu, Csaki:2017spo} we expect the twin neutrino decay  process $\nu' \to \nu e^+ e^-$ is highly suppressed.
Furthermore, it is also a rare process for the twin neutrino production at the hadron collider such as the process $pp \to W^\pm \to e^\pm \nu'$.

In the fraternal twin Higgs model, however,
although the mixing angle $\theta_{\nu\nu'}$ is suppressed with the expression
$\theta_{\nu\nu'} \simeq \lambda_{ik} \lambda^*_{kj}\frac{v f}{M_k^2}$,
interesting phenomenology may arise in this case: the neutrino and sterile neutrino oscillation in the $3+1$ pattern.
The twin neutrino with mass only a few times larger than the mass of neutrinos
will cause the $3+1$ neutrino oscillation, and thus the twin neutrino plays the role of the light sterile neutrino.

\section{Conclusion}\label{sec:Con}

In this work we have investigated the common explanation on
the little hierarchy problem,  the origin of matter-antimatter asymmetry, the tiny neutrino masses,
the nature of dark matter, and the cosmic coincidence that
the amount of dark matter and visible matter in the Universe are of the same order,
which is referred to as the ``twin cogenesis'' mechanism.

We consider the twin cogenesis mechanism within the mirror twin Higgs models which solves the little hierarchy problem.
Three heavy right-handed Majorana neutrinos are introduced to the SM sector and the twin sector respectively
to explain the tiny neutrino masses through the seesaw mechanism.
The mass mixing of the six heavy Majorana fields allow their mass eigenstates coupling to both sectors,
generating the lepton asymmetry and the twin lepton asymmetry simultaneously.
After the CP-violating out-of-equilibrium decay of the heavy Majorana fields,
there is no interaction exchanging particle asymmetries between the SM and the twin sector.
Thus the asymmetries generated via the decay of the heavy Majorana fields then freeze inside each sector.
The (twin) lepton asymmetry subsequently transfers to the (twin) baryonic sector via (twin) sphaleron processes.
Twin light states as well as the symmetric component in the twin sector will annihilate to SM electron-positron pairs mediated by the massive dark photon with Stueckelberg mass of $\mathcal{O}(10)$~MeV or higher,
mixed with the SM hypercharge via gauge kinetic terms.
In this way, the lightest (asymmetric) twin baryons
are left to be the dark matter candidates.

The twin cogenesis mechanism applies to any viable twin Higgs model
without an explicit $\mathbb{Z}_2$ breaking in the leptonic sector.
We comment in general on the dark radiation problem for mirror twin Higgs models.
We illustrate twin cogenesis
using a newly proposed neutrino-philic twin two Higgs doublet model,
where the twin neutrino masses are lifted without breaking the $\mathbb{Z}_2$ symmetry.
Twin protons and twin neutrons with masses around $5.5$~GeV as dark matter candidates
explain naturally the amount of dark matter and visible matter in the Universe are of the same order.
We also illustrate twin cogenesis in the fraternal Higgs setup,
in which the dark matter candidate is the twin bottom bound state $\Omega^\prime_{b^\prime b^\prime b^\prime}$
with mass $\gtrsim 14$~GeV.

Finally we consider various possible signatures and predictions of twin cogenesis.
The first prediction comes from the Higgs couplings,
such as the Higgs coupling to the $Z$ pair,
deviated from the SM prediction around 4\%, which can be verified by future HL-LHC data.
With all existing experimental constraints such as $\mathcal{O}(10)$~MeV dark photon searches,
using the upper bound set on the kinetic mixing parameter $\epsilon \sim 10^{-9}$,
the spin-independent cross-section of the 5.5~GeV twin baryon dark matter
scattering off nucleon is estimated to be $10^{-47}~{\rm cm}^2$,
which is just on the edge of the sensitivity of the XENONnT experiment.
Thus improved dark matter direct detection experiments in the future
with better sensitivities have the possibility
to test the twin cogenesis mechanism with 5.5~GeV twin baryons as the dark matter.
\\\\
\noindent
\textbf{Acknowledgments: }
WZF was supported in part by the National Natural Science Foundation of China under Grant No.\ 11905158 and No.\ 11935009,
and Natural Science Foundation of Tianjin City under Grant No. 20JCQNJC02030.
JHY was supported in part by the National Science Foundation of China under Grants No.\ 11875003.

\bibliographystyle{JHEP}
\bibliography{twincoge}

\end{document}